\documentclass[sigconf]{acmart}  

\setcopyright{acmcopyright}
\copyrightyear{2025}
\acmYear{2025}

\acmConference[Conditionally Accepted]{Conditionally Accepted to ACM Symposium on Spatial User Interaction (SUI 2025)}{Nov. 10--11, 2025}{Montreal, QC, CA}

\usepackage{booktabs} 
\usepackage{graphicx} 
\RequirePackage[l2tabu, orthodox]{nag} 
\usepackage{microtype} 
\usepackage[utf8]{inputenc} 
\usepackage{float}
\usepackage{placeins}
\usepackage{colortbl}
\usepackage{tikz}
\usepackage{setspace} 

\usepackage[subtle]{savetrees} 
\usepackage{array}

\graphicspath{{figures/}} 
\definecolor{softblue1}{RGB}{220, 230, 245}  
\definecolor{softblue2}{RGB}{195, 210, 235}  
\definecolor{softblue3}{RGB}{165, 190, 225}  
\definecolor{softblue0}{RGB}{235, 240, 250}
\definecolor{faintgray}{RGB}{245, 245, 245}
\definecolor{softblue2_5}{RGB}{180, 200, 230}
\definecolor{softblue1_5}{RGB}{207, 220, 240}
\definecolor{softblue0_5}{RGB}{227, 235, 247}
\newcommand{\smallplus}{\fontsize{8.2pt}{9.8pt}\selectfont}
\makeatletter

\patchcmd{\section}{.3\baselineskip}{.20\baselineskip}{}{}
\patchcmd{\subsection}{.24\baselineskip}{.14\baselineskip}{}{}
\patchcmd{\subsubsection}{.25\baselineskip}{.12\baselineskip}{}{}
\makeatother
\begin{document}

\title{Cam-2-Cam: Exploring the Design
Space of Dual-Camera Interactions for Smartphone-based Augmented Reality}

\author{Brandon Woodard}
\affiliation{%
  \institution{Brown University}
  \country{United States}
}
\email{brandon_woodard@brown.edu}

\author{Melvin He}
\affiliation{%
  \institution{Brown University}
  \country{United States}
}
\email{melvin_he@brown.edu}

\author{Mose Sakashita}
\affiliation{%
  \institution{Fujitsu Research}
  \country{United States}
}
\email{msakashita@fujitsu.com}

\author{Jing Qian}
\affiliation{%
  \institution{New York University}
  \country{United States}
}
\email{jq2267@nyu.edu}

\author{Zainab Iftikhar}
\affiliation{%
  \institution{Brown University}
  \country{United States}
}
\email{zainab_iftikhar@brown.edu}

\author{Joseph LaViola Jr.}
\affiliation{%
  \institution{University of Central Florida}
  \country{United States}
}
\email{jlaviola@ucf.edu}

\renewcommand{\shortauthors}{Woodard, et al.}

\begin{abstract}
Off-the-shelf smartphone-based AR systems typically use a single front-facing or rear-facing camera, which restricts user interactions to a narrow field of view and small screen size, thus reducing their practicality. We present \textit{Cam-2-Cam}, an interaction concept implemented in three smartphone-based AR applications with interactions that span both cameras.  Results from our qualitative analysis conducted on 30 participants presented two major design lessons that explore the interaction space of smartphone AR while maintaining critical AR interface attributes like embodiment and immersion: (1) \textit{Balancing Contextual Relevance and Feedback Quality} serves to outline a delicate balance between implementing familiar interactions people do in the real world and the quality of multimodal AR responses and (2) \textit{Preventing Disorientation using Simultaneous Capture and Alternating Cameras} which details how to prevent disorientation during AR interactions using the two distinct camera techniques we implemented in the paper. Additionally, we consider observed user assumptions or natural tendencies to inform future implementations of dual-camera setups for smartphone-based AR. We envision our design lessons as an initial pioneering step toward expanding the interaction space of smartphone-based AR, potentially driving broader adoption and overcoming limitations of single-camera AR.
\end{abstract}




\maketitle
\begin{figure}[H]
    \centering
    \includegraphics[width=\linewidth]{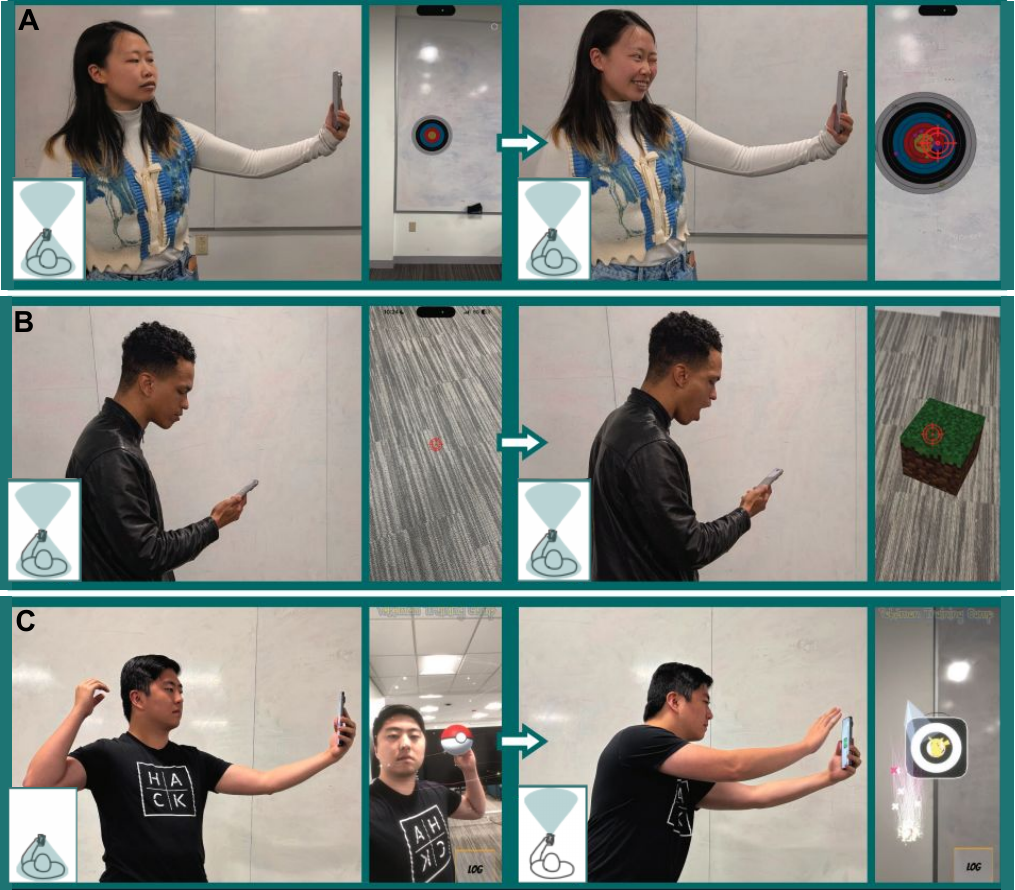}
    \caption{Cam-2-Cam leverages both rear-facing camera and front-facing camera, traditionally used independently, to extend the interaction space in smartphone-based AR applications (Left). We explore the design space through three unique applications–A.\textit{Face TriggAR} (wink gesture), B.\textit{ Mouth Craft} (mouth-open gesture), and C.\textit{ Mirror ThrowAR} (free-hand manipulation gestures) (Right). }
   
    \label{fig:Teaser_SUI}
\end{figure}

\section{Introduction}

Designing dynamic interactions for smartphone-based AR has traditionally been constrained by single-camera configurations typical in mobile devices, which either solely rely on touch-based input (e.g.,Pokemon Go \cite{niantic2024pokemongo}) or only realize the front-facing selfie camera for AR content (e.g.,Snapchat facial filters \cite{snap2024snapchat}) \cite{kim2016touch, brasier2021ar}.  Other AR form factors such as HMDs have used multiple cameras to extend the interaction space, typically allowing for more of the users' physical gestures to be captured and to expand their AR environment where users can have more \textbf{\textit{immersive}} experiences, a common goal for AR interfaces. In contrast, smartphone AR usually only leverages either the front-facing or the rear-facing camera and not both in tandem, which constrains the AR environment for smartphone AR \cite{qian2019portal,qian2020portalware,ma2023focalpoint,zhu2020bishare}. For this work, we adopt the definition of immersion presented by Agrawal et al., which states that immersion is \textit{``A state of deep mental involvement in which the individual may experience 
disassociation from the awareness of the physical world due to a shift in their attentional state."} \cite{agrawal2019defining}.
\label{intro}

To use both cameras on the smartphone meaningfully, we delegate the front-facing camera to capture a touchless physical gesture since it already faces the user, and we render the virtual reaction in the rear camera feed, situating elements in front of the user along their line of sight. This approach builds on how humans would naturally interact with objects placed in front of us in a physical space \cite{loorak2019hand, brasier2021ar, qian2020modality,qian2019portal,grubert2016glasshands}.  We aim to investigate the interaction possibilities that can motivate research opportunities centered on circumventing the limited field  

of view (FOV) and small screen size, which can hinder the AR interaction experience on smartphones \cite{qian2020modality,grubert2016glasshands,zhu2020bishare}.

To this end, we present the \textit{Cam-2-Cam}  interaction concept, implemented in three dual-camera AR applications that utilize different camera input modalities (face or hand gestures) and camera configurations (alternating or simultaneous capture) to capture the breadth of the dual-camera interaction space. Each application provides a unique example of interaction types feasible with dual-camera systems: \textit{Face TriggAR} uses a wink gesture (i.e. one-eye-closed) to simulate aiming and shooting mechanics, \textit{Mouth Craft} enables block placement with mouth gestures, and \textit{Mirror ThrowAR} integrates free-hand throwing detection across front and rear cameras (see figure \ref{fig:Teaser_SUI}). In addition to expanding on gesture-based interactions, \textit{Cam-2-Cam} draws on the concepts of responsive multimodal feedback, incorporating haptic, visual and auditory cues to enhance users' perception of control and cohesion across dual-camera setups, informed by past work on multimodal feedback integration in AR systems \cite{brasier2021ar,qian2019portal,alem2011handsonvideo,aliprantis2019natural}.

Our primary goals of this study are to construct design lessons from participants' responses that can connect the two physical spaces captured by the front and rear-facing cameras, creating a \textit{cohesive} or unified AR interaction space. We define \textit{cohesive} throughout this paper as a sense of connection between states or scenes, which is analogous to research centered on virtual transitions in XR and camera techniques in cinema \cite{pointecker2022bridging,mayer2024crossing,auda2023actuality,roo2017one, sakashita2023vroxy,tseng2013cohesion}. Due to the novelty of dual-camera setups for smartphone-based AR interactions, our study takes the form of an exploratory experiment where we use varying input modalities and camera techniques to cover the breadth of the space, highlight key design lessons in expanding the smartphone AR interaction space, and the significance of a larger interaction space concerning the goals of AR research.

We conducted the exploratory user study with 30 participants who partake in the \textit{Cam-2-Cam} experience where users try all three applications in order of increasing gesture complexity to explore the implications of dual-camera interactions, identifying key themes through qualitative analysis by answering the following research questions:

\begin{itemize}
    \item \textbf{RQ1}: How should we design the interplay between the front and rear cameras to create a cohesive interaction space?
    \item \textbf{RQ2}: What behaviors do users naturally adopt that are not part of the interface or experiment instructions that are helpful for the dual-camera setup? 
\end{itemize}

Our qualitative analysis revealed insights into what UI techniques can extend the interaction space for smartphone AR and how the larger interaction space may lead to more user engagement, leading to wider adoption of smartphone AR. We identified three primary themes to create effective dual-camera AR interactions, which we detail below.
 
\begin{enumerate}
    \item Dual-Camera Interplay: Emphasizes the importance of creating AR interactions with real-world contexts that can translate to the form factor of dual-camera, smartphone AR.
    \item Reinforcing Dual-Camera Interactions: Virtual reactions that deliver high-quality feedback to users, creating a sense of connection between their physical space, as captured by the front-facing camera, and the AR elements displayed in the rear-camera feed.
    \item Cohesive Interaction Across Cameras: A sense of fluid interactions between cameras.
\end{enumerate}

Our themes inform design lessons centered on how an expanded interaction space may make the AR experience more engaging for users and how to improve the dual-camera setup in the future. We made sure our three applications would cover the breadth of the space by implementing varying input modalities and camera techniques, including (1) camera switching, (2) simultaneous capture, (3) free-hand manipulation, and (4) facial gestures as input. The rationale for each design choice is detailed further in Section \ref{sec:designsection}. 


\section{Related Work}
\label{sec:related work}

This section summarizes prior literature that motivates the exploration of dual-camera smartphone AR. We describe smartphone AR touchless interactions with a single camera, work that aims to extend the interaction space of smartphones using additional hardware, and previous use of both cameras for AR. 
    
\subsection{Smartphone AR Interactions with Touchless Gestural Input}

Touchless gestural input has been widely explored in smartphone-based AR to provide a user experience that is natural to users based on familiar actions they'd perform in real life. For instance, Brasier et al. presented a study to offload parts of the on-screen GUI to allow users to interact with hand gestures, allowing more screen content to be visible \cite{brasier2021ar}. Qian et al. expanded on this and designed a free-hand manipulation system to manipulate AR objects using the rear-facing camera on a smartphone where they found through an iterative design process that haptic, visual, and sound cues assisted users in their AR tasks \cite{qian2019portal}. Additionally, Loorak et al.  presents hand and face interactions to interact with the on-screen UI, demonstrating users' preference for this touchless interaction versus a touch-based interface when taking selfie photos \cite{loorak2019hand}. The literature above illustrates the potential of touchless interactions, but they also share drawbacks they encountered when designing an AR experience for a single camera, as well as limited screen space with a narrow field of view.

\subsection{Expanding the Interaction Space of Smartphones}

Expanding the camera view of smartphones has been explored to expand the user interaction space and environmental awareness. \cite{grubert2016glasshands} introduce GlassHands, which uses reflections from sunglasses to enable interaction around unmodified mobile devices. Yeo et al. present OmniSense, a smartphone system enhanced with an omnidirectional camera to provide better around-the-body awareness \cite{yeo2023omnisense}. These approaches expand the field of view of mobile devices, allowing for more immersive, spatially demanding interactions -- these findings share common insights with literature centered on expanding the input space of head-mounted displays with multi-camera configurations \cite{zhu2020bishare, surale2019tabletinvr, arora2018symbiosissketch, sakashita2023vroxy}. \textit{Cam-2-Cam} similarly seeks to broaden the interaction capabilities of smartphones, but focuses on using the existing front and rear cameras to capture different types of input and output without additional hardware.

\subsection{Dual-Camera Applications in Mobile Tracking and Smartphone AR}

Prior work has demonstrated how smartphone cameras working synchronously can enhance real-world conditions experiences by providing additional sensing and improving environmental understanding. For instance, Babic et al. introduced SIMO, a system that uses the front camera for face and head tracking and the rear camera to estimate depth from a distant display to approximate body tracking \cite{babic2020simo}. Nagai et al. presented HandyGaze, which uses gaze and head tracking via the front camera combined with the rear camera and pre-scanned 3D model of the environment to localize users' gaze direction to identify artwork and present relevant information on a website in a mobile browser \cite{nagai2022handygaze}. 

Zhao et al. used both the front-facing and rear-facing cameras to capture more accurate ambient lighting, enabling more realistic shaders and textures for AR try-on applications \cite{zhao2023multi}. To the best of our knowledge, this is the only prior work that explicitly employs a dual-camera configuration for augmented reality on smartphones. However, while Zhao et al. focus on visual fidelity, the role of dual-camera setups for smartphone AR interaction techniques remains largely unexplored. Our work addresses this gap by investigating the interaction design potential of dual-camera configurations, extending the interaction space by delegating specific functions to each camera, without modifying the smartphone’s form factor. 

Rather than conducting premature comparisons with single-camera AR interfaces, we first establish a foundation for the nuanced parameters, such as gesture type, transition mechanics, separate or shared input and AR rendering, and underlying user values that may lead to future comparison studies. To support this, we implemented varying input modalities (face and hand gestures) and camera techniques (simultaneous capture and alternating views) to surface design insights specific to dual-camera smartphone AR.

 or shared input and content camera capture, and user values.

\label{sec:designsection}

\section{Cam-2-Cam Design}
\textit{Face TriggAR}, \textit{Mouth Craft}, and \textit{Mirror ThrowAR} serve as design probes for the \textit{Cam-2-Cam} interaction concept. Each application expands the smartphone AR input space by delegating the front-facing feed gestural input and the rear camera feed to situate AR content. We delegated the rear-facing camera to display the resulting AR elements from users' gestural actions captured by the front-facing camera because AR research and mainstream AR applications are primarily concerned with interacting with the physical world in front of the user (i.e., along the line of sight of the user). Additionally, our implemented applications incorporate audio, visual, and haptic feedback to enhance user engagement and provide immediate multimodal confirmation of actions. This design approach aligns with prior mobile AR research \cite{ma2023focalpoint,qian2019portal,qian2020portalware,wu2024enable} emphasizing the benefits of multimodal feedback and touchless gestures in improving screen occlusion caused by limited screen real estate and enhancing user perceived control in smartphone-based AR applications.

\subsection{Face Gesture Task Preferences}
To inform gesture-to-task mappings in \textit{Cam-2-Cam}, we conducted a within-subjects preliminary study (N = 10) evaluating six facial gestures-wink, mouth open, smile, tongue out, head nod, and head tilt—as application input triggers across two dual-camera AR tasks: block object placement and projectile shooting.

\begin{figure}
    \centering
    \includegraphics[width=0.85\linewidth]{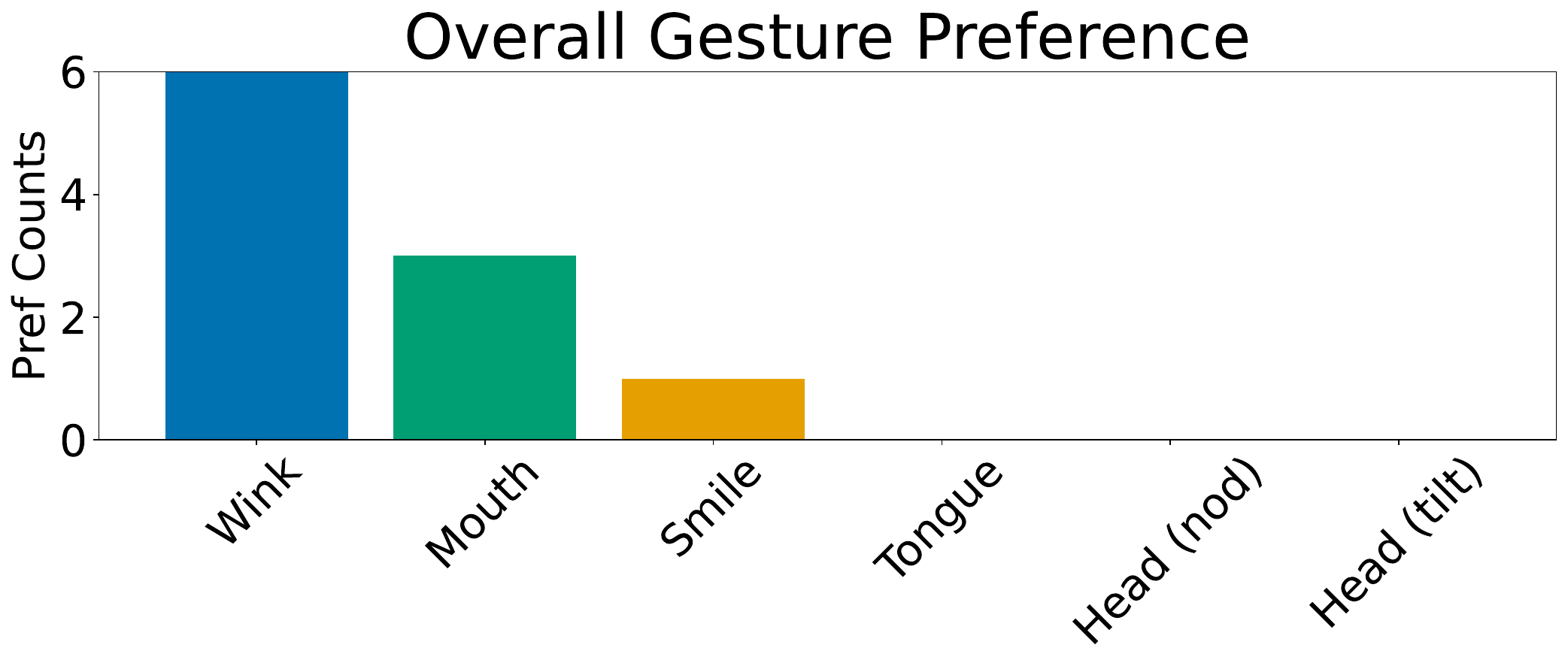}
    \caption{\textit{Bar chart shows the number of participants selecting each gesture as their preferred dual-camera AR facial input after trying out projectile shooting and block object placement prototypes.} }
    \label{fig:Prelim-Data-Counts}

\end{figure}

Participants took into account general comfort and performance when selecting their preferred gesture for each task. To avoid disrupting targeting alignment and spatial orientation in dual-camera interactions for full-head-motion gesture triggers, users nodded when placing blocks and tilted their head sidewise when shooting.

The results of Figure \ref{fig:Prelim-Data-Counts} show that wink and mouth open received all but one AR facial gesture favorites, with mouth open receiving the most counts for placing blocks and wink receiving the most for shooting projectiles. Saturated preferences for wink and mouth open from our preliminary study informed our design choice for facial gestures used in \textit{Face TriggAR} and \textit{Mouth Craft} respectively.

\subsection{Mobile Platform Considerations}

We implemented both \textit{Face TriggAR} and \textit{Mouth Craft} on iOS (iPhone 15) since simultaneous dual-camera capture with face tracking is currently not officially supported on Android.

To enable hand gestures for free-hand manipulation of AR content we used the \textit{Portal-ble} free-hand AR library due to its efficacy in manipulating AR objects as shown in several studies \cite{qian2019portal,qian2020portalware,ma2023focalpoint}. The \textit{Portal-ble} library is only supported on Android, so we implemented \textit{Mirror ThrowAR} application on Android (using the latest Google Pixel 9) 
 rather than iOS. 

 Additionally, current mobile operating systems (iOS or Android) do not allow for hand tracking combined with simultaneous capture, so we employed an alternating camera technique to capture the hand gesture with the front-facing camera and render the resulting AR content in the rear-camera feed; this approach allows us to circumvent these limitations.

\subsection{Face TriggAR}
    
\textit{Face TriggAR} is developed with \textit{Swift}, using frameworks for AR mobile scenes (\textit{ARKit} and \textit{RealityKit} \cite{apple2024arkitrealitykit}), built for iOS smartphones and later deployed on the iPhone 15. It uses world, motion, and scene tracking with a single multi-camera capture session with simultaneous front and rear-cameras for a paintball target practice simulation.

\subsubsection{Scene Recognition and Target Generation:}
The AR scene rendered in the rear-camera view handles scene recognition, paintball shooting, ricochet with gravity projectile physics, target anchoring in world space, collision bounding boxes, and procedurally generated paint splats upon paintball collision with 3D shrapnel primitives scattering. The front camera enables eye blink motion tracking with infrared depth map (\textit{TrueDepth} sensor on iOS) facial detection with which we use to continuously scan for changes in user wink input that triggers a scoping zoom with overlayed red crosshair, Western duel sound effects, and a projectile stream of paintball shots.

The system constantly scans via the rear camera for an anchor image (e.g., a QR code) to place an archery-like target while also scanning for potential eye-blink physical gestures via the front-facing camera. 
Users first pick up the smartphone in one hand and walk around to scan for a designated anchor image to fix the target onto a wall or other flat surface. They can also generate a fixed distance target in front of them manually without an anchor image.

\begin{figure}
    \centering
    \includegraphics[width=0.95\linewidth]{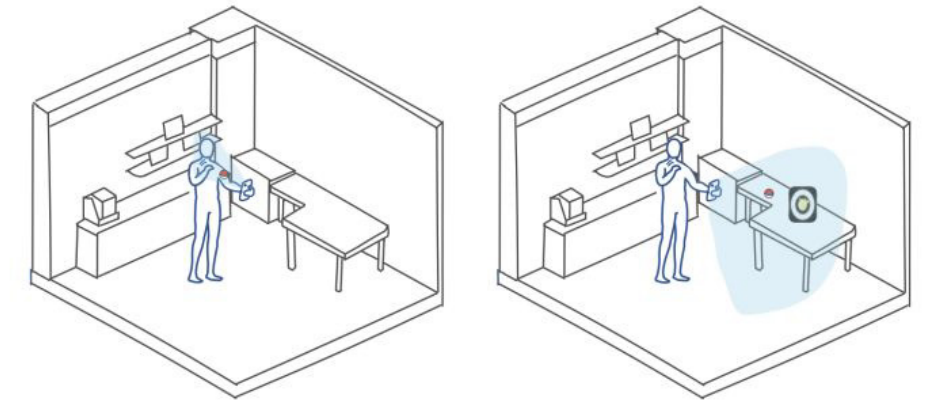}
    \caption{\textit{Face TriggAR: } (1) Displays a zoomed-in view of the user performing the wink gesture registered by the front-facing camera. (2) Shows the resulting projectiles (virtual paintballs) being shot along the user's line of sight and visualized in the rear camera's view.}
    \label{fig:TargetGenShoot}
\end{figure}

Once the user creates the target, a professional-styled archery target surface (rendered as a thin cylinder) appears, see figure \ref{fig:TargetGenShoot}. Once the target is generated after the QR code is recognized, the target is layered flush over the anchor, sharing its surface angle. 

\subsubsection{Triggering Projectile Launches:}

When a wink is detected by the mobile device's front camera, the app zooms into the scene with a crosshair display for aiming, playing Western duel audio for dramatic effect, and continuously launching virtual projectiles every half second. Our system addresses blinks (incidental moments where both eyes are shut for less than a half second) that occur while holding a wink and adequately ignores them. The app returns to its non-shooting stagnant state when the user stops holding a wink within front camera view.

The launch direction of the small metallic blue sphere projectile is derived from the camera's forward angle, with gravity simulated. Upon target impact, haptic feedback reinforces shots, and procedurally generated 3D shards of varying colors and orientations appear on-screen momentarily, with a permanent splatter left on the target.

\subsection{Mouth Craft}

\textit{Mouth Craft} is an AR application designed for the iPhone 15 that uses \textit{ARKit} and \textit{RealityKit} frameworks with \textit{Swift} to create a touchless interaction system driven by mouth gestures for object placement. 
Users can build virtual 3D structures by placing Minecraft-styled \cite{mojang2024minecraft} grass blocks into their scene environment with the front-facing camera continuously tracking facial expressions. This design allows users to act in real-time without transition delays, making the AR environment feel more immediate and responsive. When a user opens their mouth, the blocks are situated in the AR scene and rendered to the rear camera's feed.

\subsubsection{Mouth Input:}

\begin{figure}
    \centering
    \includegraphics[width=0.95\linewidth]{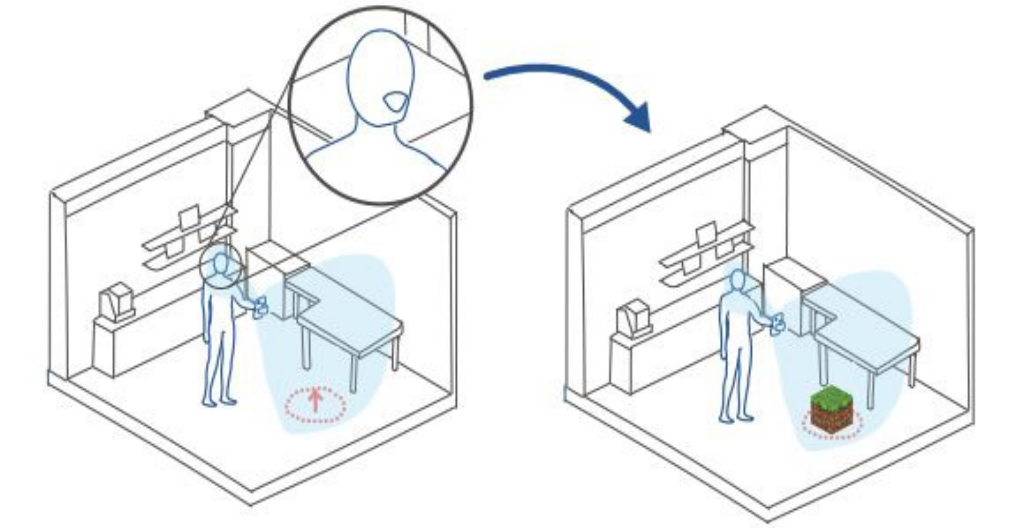}
    \caption{
        \textit{Mouth Craft: } Starting from left to right, we demonstrate the (1) user performing the `mouth open' gesture that is captured within the FOV of the front-facing camera and a detected plane is captured within the FOV of the rear-facing camera. (2) A block is placed on the detected plane and situated in the physical space captured by the rear-facing camera.
    }
    \label{fig:mouth-craft}
\end{figure}

Users can place a block by simply opening their mouth as shown in figure \ref{fig:mouth-craft}. The system captures the user's facial expressions, detects mouth-opening gestures, and uses facial recognition algorithms with an infrared depth map to monitor the degree to which the user's mouth is open. 
When the system detects that the user's mouth has opened
, a virtual cube is stacked into the AR scene. 

\subsubsection{Block Scene Placement:}

Once the user opens their mouth within front camera view, a block is situated in the physical space captured by the rear-facing camera. The cube's position is determined by the direction of the phone, and when the user opens their mouth, a forward-facing vector is projected along the direction of the rear camera from the center of the phone screen; then the block is placed in the physical space captured by the rear-facing camera where horizontal and vertical planes are detected by the LiDAR sensor on the iPhone. If other blocks are intersected by the vector when making the gesture, a new block will be placed on top of or flush against the previous one. Additionally, a crosshair is centered on the phone screen to assist users in placement fidelity.


When a block is successfully placed, users receive haptic feedback in the form of a slight vibration, and an auditory cue, such as a `pop' sound, is played. This combination of tactile and auditory feedback reinforces spatial interaction. Additionally, the system enforces a brief cooldown period of half a second between consecutive block placements to ensure that gesture triggers are intentional.

\subsection{Mirror ThrowAR}

\subsubsection{Throwing Implementation:} \textit{Mirror ThrowAR} builds upon an open-source smartphone AR system, Portal-ble, to enable free-hand throwing detection \cite{qian2019portal}. The system uses the Google Pixel 9's front-facing and rear-facing camera to track the user's spatial hand positions and convert them to 3D coordinates in the AR application in real time. These coordinates are used to detect overhand throwing gestures for our experiment.

We consider a three-dimensional Cartesian system, where $x$ is the horizontal axis, $y$ is the vertical axis relative to the virtual target generated on the wall, and $z$-axis is perpendicular to both $x$ and $y$, pointing outwards away from the cameras making the $z$-axis the forward axis in respect to the target location. Given these three axes, we can define three perpendicular planes: the $x$-$y$ plane, which describes the wall where the target is anchored; the $y$-$z$ plane that defines the vertical space between the participant and the target; and the $x$-$z$ plane, which identifies the horizontal space between the participant and the wall (see equation \ref{equation1}). 
We used ballistic motion calculations adapted from \cite{chudinov2014approximate} :
\begin{equation}
\label{equation1}
 \Theta_i = \arctan \frac{\left ( S_i^{^{2}} +  \sqrt{S_i^{4} - G \left ( Gx^{2} + 2S_i^{2}\right )}\right )}{G}
\end{equation}
where the release angle ${\vec{\Theta}}$ is affected by the release velocity $\vec{S}$, gravity $G$,  The subscript $i$ highlights the element-wise operation

\noindent performed for $\vec{S},\vec{\Theta} \in \mathbb{R}^3$ since our application operates in a 3-dimensional space.

\begin{figure}
    \centering
    \includegraphics[width=0.95\linewidth]{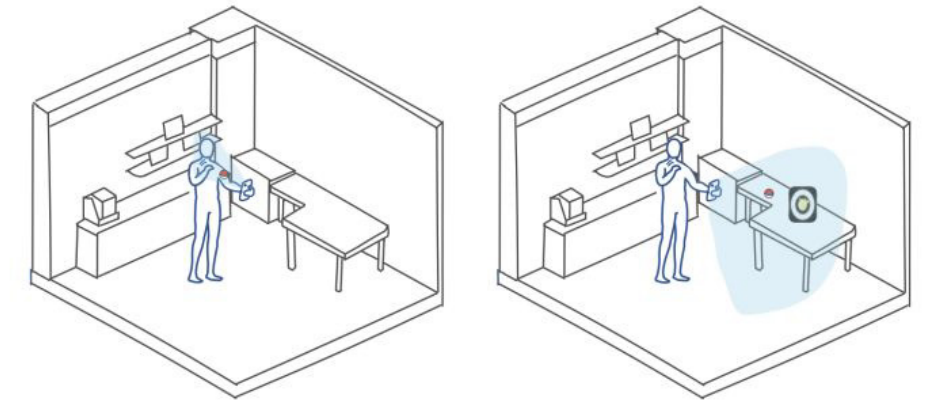}
    \caption{\textit{Mirror ThrowAR} users hold their hand out in a pinch or ball grabbing gesture to generate a Pok\'e Ball fixed to their hand visible on their smartphone. Then, they throw the ball toward their smartphone into the rear camera AR scene to try to hit the Pikachu target.}
    \label{fig:mirror-throwAR}
\end{figure}

The user performs a throw gesture toward the front-facing camera (see figures, \ref{fig:Teaser_SUI} \& \ref{fig:mirror-throwAR}), and once this action is registered, the trajectory trail continues and is displayed within the rear camera feed—creating the interaction through alternating camera views. Because the dual-camera interaction in Mirror-ThrowAR is sequential, this results in a slight latency of approximately 450 ms (0.45 seconds).

\section{Formative Evaluation}



\subsection{Participants}
We recruited participants with digital signage and email lists for on-campus graduate students at a local University. All participants were compensated \$15 and signed a consent form before proceeding with the experiment. 

Our recruitment goal was 30 participants as this is often sufficient to reach saturation of themes in qualitative experiments, and a higher number typically results in more nuanced insights \cite{guest2006many,iftikhar2023together}. The recruited participants included 12 females and 18 males between the ages of 18 and 60 years old (\(\mu = 26.97\), \(\sigma = 9.41\)). Participants also provided their experience with smartphone AR during our semi-structured interview and rated on a 5-point Likert scale (1 = No Experience, 5 = Very Experienced). The resulting levels of experience can be summarized as (\(\mu = 1.93\), \(\sigma = 0.89\)). 

\subsection{Study Protocol}
This study takes the form of a dual-camera AR experience where participants try the three developed applications (1) \textit{Face TriggAR}, (2) \textit{Mouth Craft}, and (3) \textit{Mirror ThrowAR} with order increasing by task physical complexity. This study takes place within a 3 x 4 meter space with a table and couch present so users can situate AR objects on physical objects typically found indoors. Participants used \textit{Face TriggAR} and \textit{Mouth Craft} for the duration of two minutes and threw 10 projectiles with \textit{Mirror ThrowAR}–values determined based on observed fatigue in preliminary experiment. 

Each application session began with experimenters demonstrating the functions of the applications until participants felt comfortable trying on their own without training and starting the trial. In \textit{Face TriggAR}, participants first scanned a QR code anchor image to generate a fixed AR target, spending the initial moments aiming at this anchored target. Afterwards, they could manually reset the target location such as on a table or coach, allowing them to explore aiming from various angles and even move behind the target to observe it from different perspectives. Using a wink gesture to aim and shoot at each target, participants had two minutes to interact with the scenes. 

For the \textit{Mouth Craft} application, participants were instructed to ``build a fortress or structure'' by placing virtual blocks within the AR environment using mouth gestures. They explored positioning blocks on surfaces like tables, couch, walls, and floors where a subtle popping audio feedback and a brief haptic cue followed each placement. Participants moved freely within the space to test block placement from various angles for the next two minutes.

In \textit{Mirror ThrowAR}, participants completed a sequence of 10 Pok\'e Ball throws into the AR scene. For each throw, participants began by holding the device in front-facing mode, capturing the throwing gesture through the front camera. As they released the Pok\'e Ball, the application performed a sequential camera switch to the rear camera, allowing participants to view the projectile’s trajectory through the AR space.
Participants were instructed to aim each throw at a virtual target, receiving visual feedback in the form of motion trails and collision effects, along with accompanying haptic and audio cues at both release and impact. 
 
\subsection{Data Analysis}
Experimenters conduct semi-structured interviews with the following qualitative questionnaire based on the AR Design Heuristics presented by Endsley et al. and tailored to fit our research questions displayed in table \ref{tab:survey-questions} \cite{endsley2017augmented}. The first and third authors carried out the thematic analysis by initially performing an open coding procedure followed by axial coding independently. Afterward, they reconciled differences in the final code groups and formed themes as seen in table \ref{tab:thematic_table}. 

\begin{table}[h!]
\centering
\renewcommand{\arraystretch}{1.0}
\small
\begin{tabular}{|p{0.45\columnwidth}|p{0.5\columnwidth}|}
\hline
\textbf{Survey Questions} & \textbf{AR Design Heuristics \cite{endsley2017augmented}} \\ \hline

How did your physical actions influence virtual objects? & 
\begin{minipage}[t]{\linewidth}
\begin{itemize}
    \item Form communicates function
    \item Fit user's physical abilities
    
\end{itemize}
\end{minipage} \\ \hline

How did translation between gestures and virtual reaction feel? &
\begin{minipage}[t]{\linewidth}
\begin{itemize}

    \item Alignment physical-virtual
    \item Adaptation to user motion
    
\end{itemize}
\end{minipage} \\ \hline
Moments of disorientation or disruptions (selfie \& rear camera)? &
\begin{minipage}[t]{\linewidth}
\begin{itemize}

    \item Minimize distraction
    \item Accessibility off-screen objects
    
\end{itemize}
\end{minipage} \\ \hline

How much did physical actions make virtual objects feel present? &
\begin{minipage}[t]{\linewidth}
\begin{itemize}

    \item Fit user environment/task
    \item Fit perceptual abilities
    \\
\end{itemize}
\end{minipage} \\ \hline
Which experience did you prefer and why? &
\begin{minipage}{\linewidth}
\begin{itemize}
    \item Subjective feedback
\end{itemize}
\end{minipage} \\ \hline

\end{tabular}
\caption{Survey Questions participants answered next to corresponding AR Design Heuristics for dynamic interactions from literature that informed our post-study questionnaire. In addition to this, participants were asked if they could think of application scenarios that can leverage dual-camera mobile AR and demographic questions.}
\label{tab:survey-questions}
\end{table}

\section{Results}
This section presents the findings of our thematic analysis in order to outline design lessons for our dual-camera AR setup as seen in table \ref{tab:thematic_table}.


\subsection{Context-Aware Dual-Camera Interplay}
We begin by detailing codes informed by users' responses, highlighting the importance of creating AR experiences with familiar concepts that can translate into a dual-camera setup. Additionally, we highlight contrasting responses in which users sometimes accept contextual misalignment when interpreting the AR experience as surreal.

\subsubsection{Contextually Aligned Front-facing Camera Input with Virtual Reactions Shown in Rear Camera:}
\label{tab: contextual-alignment}
Participants ({N = 16, 53\%}) shared positive sentiments about the applications and related them to what they've experienced in the real world or video game contexts. They enjoyed \textit{Face TriggAR}'s `wink-to-zoom' and holding out the phone in a way that mimics shooting in the real world, as in how they perceive shooting characteristics. For example, P16 details why they enjoyed \textit{Face TriggAR} and explains, \textit{``I thought that squinting one eye to get a cross (crosshair) was a cool, intuitive gesture and one that I would do in a non-AR setting,''} suggesting that their gesture, captured by the front-facing camera, and visual feedback aligns with how they might perform the action in real life. Additionally, participants observed a sense of realism and associated their gestural input to actions they perform in the real world. P28 associated this realism regarding the throwing gesture in \textit{Mirror ThrowAR} captured by the front-facing camera and shown in the rear camera as \textit{``it's literally throwing. That's essentially, I feel that one was probably the most real, like the physical gesture.''} 

In contrast, for \textit{Mouth Craft}, some participants felt it caused a disconnect. P23 describes the disconnect while contrasting it with \textit{Face TriggAR}: \textit{``There's some semantic connection between closing your eyes and shooting. I couldn't build such a semantic connection between opening your mouth and actually placing Minecraft cubes.''}

This disconnect described during the experiment aligns with the first \textit{AR Design Heuristics} presented by Endsley et al. \cite{endsley2017augmented}. The observed disconnect is essential to note as to provide a cohesive interaction experience for a dual-camera set-up, users should feel a sense of connectedness on their side of the phone (front-facing camera) to the virtual reaction shown in the rear camera. Although, a contrasting attitude towards \textit{Mouth Craft} was described by P8 where they tell their positive experience with the mismatch of input and AR scene: \textit{``So that felt more surreal, but I think that kind of makes it more fun because it's kind of like `Dr. Seussy'.''} In this case, the participant referenced a children's book author, Dr. Suess, known for their otherworldly stories that aren't always aligned with real-world semantics. Therefore, P8 makes a connection between a familiar artistic form of expression to contextualize their experience with their unique subjective interpretation.

Further, concepts of contextual relevance (contextually aligned front-facing camera input and virtual reactions) are seen in some responses related to the actual form factor of the smartphone itself and the types of AR `context' the phone can provide. P21 describes the holding of the phone and closing one eye for \textit{Face TriggAR}: \textit{``The entire posing of it, too, is really nice because you're holding this phone far from you. You close an eye, and it's like if you're doing archery/your gun, you hold the tool out, so it's like you're role-playing this scenario.''} This response focuses on the form factor in the contextual relevance of a dual-camera setup, where P21 draws a connection between how they hold the phone away from their body, close one eye for aiming, and relates this to a real-world sport, archery. The takeaway in P21's sentiments is how specific AR interactions on a smartphone with a dual-camera setup should consider characteristics of the form factor if the intention is to provide users with context. 

\subsubsection{Alignment of Input Types and Camera Views:}
Several participants (N = 17, 56\%) expressed attributes of contextual relevance with the alignment of cameras or FOVs and limited screen space. P2 describes the alignment of the front-facing camera to their face, complementing the context and realism of the task: \textit{``The interaction is like a real action, and it's really helpful to do something in this way. The selfie camera can see your eyes because it's already capturing your face.''} This is complemented by P20's response to question \#2 about \textit{Face TriggAR}, \textit{``I'm looking at the target on the screen so that one translates pretty well.''} These statements may demonstrate how the front-facing camera is inherently designed for capturing one's face, and closing one eye to aim assists in creating an experience that's realistic, and proper alignment of the front-facing camera may allow users to make a stronger connection to real-world contexts. P20 seemed to have felt that alignment with the virtual target they're viewing makes sense, intuitively, to how they feel they \textit{should} be viewing the target. P24 further explains how they feel different FOVs may make more sense with different application themes: \textit{``I felt like I had to move the camera a lot to get information for where the blocks in my scene was ... I associate a tighter FOV with, like, more precision, which has worked really well for the task.''} P24 first describes difficulty with \textit{Mouth Craft} in not being able to see all of the blocks they placed within the single FOV and how they associate the same FOV used in \textit{Face TriggAR} with precision, thus a shooting game made more sense with regards to the limited front-facing and rear camera FOVs. This may implicate a consideration of the contexts of motifs or actions with respect to the FOVs in a dual-camera AR setup.

\subsection{Reinforcing Dual-Camera Reactions}
This theme concerns the quality of multimodal feedback participants noted, which influenced their overall experience. This theme captures the sentiments of participants related to the type of AR feedback participants experienced. We identified two distinct types of feedback participants felt influenced their experience: (1) Bodily actions and responsiveness and (2) Affirming Multimodal Feedback in Dual-Camera AR.

\subsubsection{Bodily Action and Responsiveness:}
Most participants (N = 22, 73\%) felt the responsiveness of the virtual reactions gave them a greater sense of \emph{connection}. P13 describes \textit{Mouth Craft} as \textit{``I feel like the fact that the as many times as you open your mouth, each time a block, would just come up, automatically...It just felt like those are kind of linked.''} This response highlighted a commonality among users when it came to the responsiveness of \textit{Mouth Craft}, and despite the misaligned semantics of the application, the responsiveness induced feelings of a `link' or connection between the physical gesture and the responsiveness for P13, thus creating a sense of cohesiveness between the front-facing camera input and rear camera output. This sense of a cohesive interaction is tied to the responsiveness affirming their actions. Participants also described their experience with the \textit{Cam-2-Cam} applications as feeling `connected' to the virtual objects rendered in the rear camera's view, and they also described a sense of `disconnect' when an application was not as responsive. 

There may also be a discrepancy when an object is unexpectedly responsive. Two participants categorized in this group, P10 and P22, also shared contrasting sentiments for one of the applications. P10 describes an unexpected response when launching projectiles in \textit{Face TriggAR}: \textit{``In the first experiment ... it wasn't fully responsive when I kept an eye shut. There was a bit of a disconnect.''} P10 shares an issue with the responsiveness that may have been due to problems with gestural recognition during their experiment. As a result, they explicitly describe, thus contrasting, the views of users who related responsiveness to a connection. This statement can highlight a contrasting side of the spectrum regarding responsiveness and creating a cohesive interaction space - balancing may be necessary for different users where reaction time across cameras is considered when designing similar dual-camera UIs. 

\begin{table*}
    \centering
    \smallplus
    \renewcommand{\arraystretch}{1.25}
    \resizebox{0.9\textwidth}{!}{%
    \begin{tabular}{|>{\raggedright\arraybackslash}p{4cm}|>{\raggedright\arraybackslash}p{4cm}|>{\raggedright\arraybackslash}p{8cm}|}
        \hline
        \textbf{Themes} & \textbf{Codes} & \textbf{Exemplar Quote} \\ \hline
        
        \cellcolor{softblue2_5}
        Context-Aware Dual-Camera Interplay \textbf{(N = 24, 80\%)} & 
        \cellcolor{softblue1_5}Contextually Aligned Front-facing Camera Input with Virtual Reactions Shown in Rear Camera \textbf{(N = 16, 53\%)} & 
    P24: ``I've seen people that close one of their eyes to aim, and for some reason, I associate that with precision. That worked really well for that piece, specifically.'' \\ \hline

         &\cellcolor{softblue1_5}Alignment of Input Types and Camera Views \textbf{(N = 17, 57\%)} & 
          P25: ``The only thing I experienced was sometimes if you had your face too close or too far away, then the camera wouldn't capture ... I guess, kind of wouldn't be in the frame.'' \\ \hline
        
        \cellcolor{softblue2_5}
        Reinforcing Dual-Camera Reactions \textbf{(N = 22, 73\%)} & 
        \cellcolor{softblue2_5}Bodily Actions and Responsiveness \textbf{(N = 22, 73\%)} & 
         P9: ``I feel like the translation was the best for the open-mouth task (\textit{Mouth Craft}); that was the most fluid. — I think I preferred the mouth one because it felt the most responsive.'' \\ \hline
         
        & \cellcolor{softblue0}Affirming Multimodal Feedback in Dual-Camera AR \textbf{(N = 10, 30\%)} & 
         P5: ``For opening your mouth and placing an object,  the sound effects make it more realistic. The sound effects sound like a person opening their mouth.'' \\ \hline
        \cellcolor{softblue1_5}
        Cohesive Interaction Across Cameras \textbf{(N = 19, 63\%)} & 
         \cellcolor{softblue1_5}Seamlessness of Camera Transitions \textbf{(N = 18, 60\%)} & 
          P25: ``I felt like there was a smooth transition for all of them, like the graphics, you can see it slowly appearing.'' \\ \hline       
        
        & \cellcolor{softblue0}Visual affirmation of physics \textbf{(N = 9, 30\%)} &  
        P28: ``I was playing with the Pok\'e Ball; I could try to aim to where I'm casting it [the Pok\'e Ball] off to in the camera frame.'' \\ \hline
\multicolumn{3}{c}{
    \raisebox{-1.5em}{
        \begin{tikzpicture}[scale=0.5]
            \shade[left color=softblue0, right color=softblue3] (0,0) rectangle (6,0.4);
            \node at (0, -0.3) {\scriptsize Low (N)};
            \node at (6, -0.3) {\scriptsize High (N)};
        \end{tikzpicture}
    }
} 
    \end{tabular}
    }
    \caption{Themes and codes with Exemplar Quotes (heat map shading indicates prevalence of category) }
    \label{tab:thematic_table}
\end{table*}

\subsubsection{Affirming Multimodal Feedback in Dual-Camera AR:}
(N=9, 30\%) participants highlight the potential importance of multi-modal affirmation aligned with the AR reaction. P21 describes how the haptic vibration affirmed their actions and observed reactions: \textit{``I have feedback from this virtual weapon, or whatever you want to call it, and I think that that made it quite convincing to me that I was like, this is, real in a sense.''} In this particular quote, the user underscores how the haptic feedback created a sense of realism by affirming their input gestures and virtual reactions. 

Participants also emphasized the alignment of the multi-feedback and how this contributed to a cohesive experience with the dual-camera interaction. P7 describes how the mouth gesture of opening your mouth makes intuitive sense with the `popping' noise: \textit{``your mouth kind of goes (pop – participant does mouth popping noise). The noise did it, too. So seeing that and hearing that, and feeling you do that was all very connected. When you go, like this (Participant demonstrates the mouth gesture) and the noise goes with it, and the visuals too.''} The affirming visuals and sounds contributed to the participant's overall sense of the dual-camera interaction space feeling connected. P5 echoed this sentiment but with a focus on the sound effects: \textit{``I think for opening your mouth and placing an object, the sound effect makes it more realistic.''}

\subsection{Cohesive Interaction Across Cameras}

Participant responses revealed how camera transitions can affect the perceived connection of their space captured by the front-facing camera and the AR elements overlaid within the rear camera view. This can lead to a fluid interaction, thus allowing more space around the phone to be usable. The following codes that may influence camera transitions uncovered in our analysis are visualizing AR physics, seamlessness, and responsiveness of alternating cameras. We define \textit{seamlessness} as participants' sentiments related to smooth or gradual changes that make intuitive sense between their gestural input and virtual response, such as a fade-in or fade-out. \textit{Responsiveness} in this context refers delay at which the two cameras take to alternate between each other.

\subsubsection{Seamlessness of camera transitions:}
Uninterrupted changes between cameras or seamlessness were described by (11, 37\%) participants. When asked about their feelings towards their physical gestures and virtual reactions (Q2), P14 describes an obvious \textit{cause-and-effect} between the front-facing camera delegated to capture their gesture and rear-facing camera intended to display the AR reactions: \textit{``In the first one (\textit{Face TriggAR}), it was pretty much seamless... When I close one eye, it does that (virtual reaction).''}. P14's sentiment seemed related and supported by P25's response: \textit{``It felt, I felt like there was, like, a smooth transition for like, all of them, like the graphics you can see it slowly, like, appearing.''}. P25 details their experience in a way similar to how seamlessness is described in camera transitions or AR/VR transitions on HMDs, where the emphasis is placed on the gradual change between states (e.g., action and reaction) \cite{pointecker2022bridging, mayer2024crossing, auda2023actuality, roo2017one,sakashita2023vroxy,tseng2013cohesion,shan2024work}. 

P11 echoes the sense of seamlessness and details specific features which may attribute themselves to a seamless experience in \textit{Mirror ThrowAR}: \textit{``after it landed on the target, it flipped back to me pretty smoothly, and the flipping from front to rear as I was throwing the ball was pretty smooth''}. P11's description is centered on the actions occurring in-game and the cadence of the cameras alternating. When a ball is thrown in \textit{Mirror ThrowAR}, there is a visible camera transition where the frame is frozen until the rear camera's feed starts. Once started, the elements from the AR scene overlay onto the rear camera feed, and the projectile and trajectory become visible; lastly, once the ball lands, the user can observe its landing location for five seconds. The application switches back to the front camera feed so a user can begin throwing again. 

In contrast, some participants also expressed disruptions or lack of seamlessness between cameras. P1 suggests addressing what they perceive as a non-seamless transition in \textit{Mirror ThrowAR} when describing any moments of disorientation (Q3): \textit{``I'm looking at myself, I couldn't see the target initially. If there is a separate line where one is streaming with the selfie camera and the other is streaming with the rear camera, that might be better.''} P1 highlights a difficulty with the alternating camera feeds when using \textit{Mirror ThrowAR} and suggests a `split screen' view as seen in multiplayer video games. P1's suggestion may allow the interaction space also to be extended visually, where two AR scenes can be rendered on both camera feeds, thus increasing the visibility of AR elements and their resulting interactions across cameras. We envision using a method similar to mobile video conferencing tools that delegate the larger feed to the area of focus and the smaller feed in the bottom right corner, forming a picture-in-picture interface for an AR scene.

\subsubsection{Visual affirmation of physics:}

(N=9, 30\%) participants expressed how they felt assisted or disoriented by recognizing visual feedback of the AR physics implemented in the \textit{Cam-2-Cam} applications. P28 described creating a mental map while using \textit{Mirror ThrowAR} between the direction of the trajectory trail and Pok\'e Ball leaving their hand versus where the ball would land on the target: \textit{``I tried to aim where I'm casting it (Pok\'e Ball \& trajectory trail shown in rear camera feed) into the camera frame. Which is why I think I did better. Once you figure out that one, you know ... okay, I need to throw it in the general area.''}. P28 used the `casting' direction of the ball within the boundaries of the phone screen (frame), rendering the selfie camera feed to where the projectile lands in the rear camera feed. Their description of a mental map between the two spaces captured by the front-facing and rear cameras may be interpreted as an affirmation of their logical connection. 

P14 describes a similar mapping enabled by visualizing physics between their physical actions and the visible trajectory of the paintball in \textit{Face TriggAR}: \textit{``Moving the camera up and down would also sort of change how the ball was being shot...  it was like this connection between what I was doing and what was in AR.''}. P14 acknowledges that the visible projectile motion intuitively made sense between their physical actions and the AR space, thereby creating a connection between themselves and the AR scene in the rear camera feed displayed on the phone. 

Not all participants were positive about the visible AR physics, and most notably, P10 shares that the AR physics they observed may not have been logical responses to their physical actions when using \textit{Mirror-ThrowAR}. P10 states, \textit{``The physics in the virtual environment were not related to the physics happening in the real world... a harder throw didn't mean a faster Pok\'e Ball getting thrown at the target.''}. P10 expresses an opposite effect based on how they interpreted their physical actions and the AR physics, possibly causing a disconnect between the two spaces, as highlighted when they felt the Pok\'e Ball's speed didn't align with what they've experienced in the real world. In particular, \textit{Mirror ThrowAR} is unique in that users start throwing with the front camera feed, where their trajectory continues into the rear camera feed. P10 describes a disconnect between when they throw into the front camera feed and the resulting projectile and trajectory once the ball is released into the rear camera feed. It's also worth noting that P10's response to Q4 centered on the connection between users' physical actions and virtual objects. This sentiment further supports how the AR physics must be meaningful between the front-facing and rear cameras, or they may cause a disconnection.



\section{Discussion}

\label{Discussion}

\subsection{Design Lessons for a Cohesive Dual-Camera Interaction Space (RQ1)}
We contribute to smartphone-based AR research that leverages dual-camera setups by capturing user impressions via our questionnaire aiming to identify design lessons related to cohesive interaction between the front-facing camera and the rear-facing camera. Our concept of a coherent dual-camera interaction space adapts the definition of cohesiveness we outline in Section \ref{intro} to focus on designing a dual-camera interface where the spaces captured by the front and rear cameras can be perceived as connected to each another. 

The following design lessons detail the interface and interaction design informed by our experiment, which may allow for a cohesive interaction space for dual-camera AR interactions, expanding the limited form factor without modifying the device's hardware. 

\subsubsection{Design Lesson -- Balancing Contextual Relevance and Feedback Quality (Themes \#1 and \#2):}

We discovered an interdependent relationship between themes \#1 and \#2, where participants could be more accepting of a lower context awareness of the themes and interaction of the application if the feedback quality implemented was engaging enough and vice versa. Therefore, we named this design lesson, \textit{Balancing Contextual Relevance and Feedback Quality}. 

\textit{Face TriggAR} and \textit{Mirror ThrowAR}'s strengths were centered on the familiarity of users with interactions in other real-world contexts. In contrast, \textit{Mouth Craft}, where users could not draw a connection from the real world of placing blocks by opening their mouths, still received praise for the feedback quality people experience through sound effects, haptic vibrations, near-immediate action response, and visual elements. This multimodal feedback allowed users' actions to feel `connected' to the blocks they were placing, thus implying a connection between the user's space and the AR space overlaid onto the rear-facing camera's feed. 

Molnar et al. relate the immediacy of feedback from AR elements and the type of feedback in response to actions to flow and immersion \cite{molnar2019augmented} . In addition, immersion in an AR context can be defined as the presence and connection to the AR elements with which users interact \cite{molnar2019augmented}. The majority of positive responses for \textit{Mouth Craft} were categorized under the theme, \textit{Reinforcing Dual-Camera Reactions}, which implies the experience of the application can be attributed to its multimodal feedback in response to users' gestures. 

P5 responses captured the lack of familiar context with \textit{Mouth Craft} but still preferred it among the three applications due to the immediacy of feedback. P5 describes \textit{Mouth Craft} as feeling, \textit{``weird because these two (mouth open gesture and placing a block) make no connection in my mind''}. However, P5 follows up to explain that \textit{Mouth Craft} was their favorite, \textit{``because it's more responsive''} and also states, \textit{``the sound effects sounds like a person opening their mouth''}. P5's responses about \textit{Mouth Craft} underscore a familiar and somewhat contradicting sentiment across multiple users where they still preferred \textit{Mouth Craft} due to the immediate feedback of the AR scene.

P15 echoes the immediacy of the feedback as benefiting the translation between their physical gestures and virtual reaction (Q2): \textit{``It felt pretty natural and consistent — when I opened my mouth, for the most part, it would place a block.''} and, \textit{``There was a connection between what I was doing with my body and what was appearing on the screen ... I noticed it most with the mouth one (\textit{Mouth Craft}) where I would do a thing, and then the tiny computer in my hand would respond to it by creating something I could see.''}. What P15 describes is a sense of connectedness or embodiment to the virtual objects, which plays a critical role in the overall immersion of an AR interface \cite{genay2021being, sakashita2023vroxy}. Additionally, despite the unnatural or unexpected virtual response, \textit{Mouth Craft} was still able to induce a `natural' feeling for P15. These sentiments could indicate a design opportunity to make more engaging smartphone AR applications by introducing a dual-camera setup where a user's space can feel connected to the space captured by the rear-facing camera. 

Responsiveness and quality feedback supplementing feelings of embodiment and immersion have been widely studied in smartphone AR and other XR form factors. In smartphone AR games, Xu et al. investigate how a sense of embodiment and connections to virtual characters or AR elements can be achieved with immediate, sound, and visual responses \cite{xu2011pre, feuchtner2017extending}. On the opposite end of the spectrum, the literature demonstrates how a lack of haptic, sound, and visual cues can cause a disconnect or reduced sense of embodiment \cite{abtahi2022beyond}. Our work extends this prior literature by demonstrating a causal relationship between the immediacy of AR feedback and multimodal sensory cues. Further, the immediacy of feedback and multimodal cues have been shown to induce feelings of connectedness in non-AR contexts \cite{iftikhar2023together}.

We believe the dual-camera setup for smartphone AR opens up a design opportunity for making the space captured by both the front-facing and rear-facing cameras be perceived as connected, thus expanding the perceived space and possibly creating a more engaging experience previously limited by the form factor. Once the gestures captured in the users' space feel connected to what is \textit{rendered on the other side of the phone} through the rear-facing camera, the embodiment the user experiences with the virtual elements may increase their presence and overall engagement, which may lead to broader adoption of smartphone AR as being a more common medium for AR experiences. Most popular smartphone AR experiences are constrained to a touch-based interaction, which is an everyday reminder that the AR scene is on the phone, whereas in \emph{immersive AR} experiences, you almost want the user to forget about the form factor involved completely \cite{weber2021get, qian2019portal}. 

Based on our findings, a research opportunity exists to compare single-camera touched-based interaction to the dual-camera setup for AR to identify more precise trade-offs about immersion. Creating experiences where smartphone-based interactions allow users to be more immersed could lead to broader adoption as a common theme with all XR form factors is a lack of success due to poor immersion induced by their interfaces \cite{genay2021being, sakashita2023vroxy}.

We suggest a design model where an application's familiarity is based on real-world context and feedback quality that can be used to balance these two attributes for future dual-camera AR setups. Our results identified a spectrum between contextual relevance and multimodal feedback, and this model can inform other smartphone AR applications with a dual-camera setup. In particular, this design model can be used to gauge whether the intended functionality of an AR application can be achieved without sacrificing immersion. The smartphone AR space needs workarounds that can bring a level of immersion closer to what is experienced by its HMD counterparts that also use multiple cameras to expand the interaction space, multimodal feedback, and gesture recognition.

\subsubsection{Design Lesson -- Preventing Disorientation using Simultaneous Capture and Alternating Cameras (Theme \#1 \& Theme \#3)):}

\textit{Preventing disorientation using simultaneous capture and alternating cameras} is focused on camera transitions and alignment that are predictable for the user. This design lesson is informed by responses falling under the themes, \textit{Context-Aware Dual-Camera Interplay} (Theme \#1) and \textit{Cohesive Interactions Across Cameras} (Theme \#3).

\textit{Face TriggAR} and \textit{Mouth Craft} had strengths regarding the seamlessness of captured gestures and virtual reactions. Still, a common drawback for participants was being unable to tell if their faces were in the frame for their gestures to be captured. This disorientation could be addressed with visual cues, such as a picture-in-picture (PiP) view that displays the user's face from the front-facing camera feed or an avatar overlaid onto the rear camera feed, similar to the `split screen' view suggested by P1. 
Similar adaptations of this type of PiP GUI design are widely used outside of AR contexts (e.g.,video calls) for self-orientation\cite{vanukuru2023dualstream}. 

\textit{Cohesive Interaction Across Cameras} captured sentiments from users related to their own body orientation with respect to the phone and virtual content which may indicate a separate cameras or scenes could be made to feel connected with the combination of transitions and visual affirmation of physics. These concepts, which are analogous to cinematic camera techniques, suggest that researchers may discover new overlaps between XR and cinematic effects, potentially enabling more effective action triggers and navigation in AR. A study focused on different visual transitions or camera responsiveness (e.g., simultaneous capture or alternation speed) could shed light on the Just Noticeable Difference (JND) for users between their physical actions and virtual transitions in a multi-camera smartphone AR setup, helping to quantify strategies for perceptibly connecting spaces.
Further, dual-camera setups with dual views may introduce new interaction techniques that circumvent the issues of disorientation experienced with single-camera smartphone AR, where users typically have to hold the phone while keeping their manipulating hand in-view, thus occluding parts of their arm. These spaces partitioned by the smartphone may cause incongruence between the user's hand captured by the rear-camera and their body. \cite{grubert2016glasshands,qian2020portalware,ma2023focalpoint}
Our work encourages future research to explore various methods of affirming user actions—such as self-viewing PiP GUIs or visual cues—that may help create cohesive virtual spaces in which users experience a sense of continuity between their actions and the XR content.

\subsection{Considerations of User Defined Strategies for Dual-Camera AR (RQ2)}
We identified strategies users developed on their own that were not part of the application interface or the experiment instructions, and these strategies provide valuable insights when designing dual-camera AR setups on smartphones. These strategies may be helpful considerations when designing smartphone AR with dual-camera AR. For example, P10 felt that head orientation was being registered with \textit{Face TriggAR}: \textit{``From what I remember, me moving my head also changed the direction of the screen. I think that implied a sense of control there--the head movement determined the way the thing is sprung out (shot out)\textit{Face TriggAR}), as opposed to a static image that plays every time.''} Similarly, P8 demonstrated to the interviewers how she tried to move her head to place objects for \textit{Mouth Craft} in the direction she wanted to, even though she was aware she needed to move her hand holding the phone to change the direction. These responses may present a design opportunity to integrate head movements into the dual-camera interface as it may be a natural response from users. Integrating an additional parameter like head rotation may also induce feelings of engagement as more physical input, especially natural ones, has been shown to lead to more engagement in AR research \cite{wu2024enable,qian2019portal}. Prior work has explored different theories within AR contexts, like Self-Determination Theory, to support users' agency or flexibility in how they interact with technology to provide a more natural driver for sustained engagement \cite{bennett2023does,tyack2017exploring}. 

Participant responses also revealed that users created mental maps connecting their space and gestures captured by the front-facing camera to the AR scene in the rear camera's feed. Most notably, P28, P8, and P12 describe very similar mental maps when using \textit{Mirror ThrowAR}, where they would make minor adjustments throughout their throws and eventually began understanding where their projectile landing location would be. The strategy required them to adjust based on where their hand was within the bounds of the screen as described by P12: ``So if a ball doesn't, let's say, show up, or like, appear as I expected it, I would see how my hand looks on screen to try to adjust it.'' Design approaches involving reducing the mental load with additional multimodal cues or a PiP view to simplify a dual-camera interface could also be considered. Such cues have suggested benefits in AR and GUIs that span spatially, allowing for more intuitive controls and enhancing feelings of embodiment \cite{latoschik2016fakemi,li2021vmirror,anderson2013youmove}.



\subsection{Application Scenarios}

Many participants shared application scenarios that can leverage smartphone AR with a dual-camera setup as showcased in figure 1 in our appendix demonstrating breadth of potential applications. For instance, P21 described a dual-camera AR museum application where closing either eye augments an art piece you're viewing: \textit{``So maybe with your like, if you have two eyes open, you're seeing version A, one eye open, you're seeing version B, when I the other eye version C.''} This augmentation could add more interactive experience for art exhibits where the subject viewing the artwork may become more engaged with the work. This application scenario could also lead to research centered on embodiment with art pieces.

The application scenarios in our participants' responses applied to different domains, such as art experiences, training, gaming and education. The variety of potential applications demonstrates an extensive design space smartphone AR with dual-camera setups may present. It is believed in scenario-based design, if there are a variety of use-cases for a particular system a multitude of interaction methods are likely possible \cite{carroll2003making,carrol1999five,o2020defining,iftikhar2021designing}. Our study is just a first step in exploring the dual-camera smartphone AR design space, but we foresee a diverse range of interaction styles and potential research venues that can leverage our qualitative data to inform future experiments. 

\section{Conclusion}
We presented the \textit{Cam-2-Cam}  interaction concept implemented in a series of dual-camera interactive smartphone AR applications to form design lessons on how to make the smartphone AR interaction space larger while achieving the goals of AR interfaces. Our takeaways consist of the two main design lessons captured from our qualitative analysis: (1) \textit{Balancing Contextual Relevance and Feedback Quality (Themes \#1 and \#2)} which details how interfaces with metaphors from the real world can be balanced with high-quality feedback to induce a sense of embodiment or immersion and (2) \textit{ Preventing Disorientation using Simultaneous Capture and Alternating Cameras (Theme \#1 and Theme \#3)} which outlines methods that should be considered in order to prevent disorientation using the dual-camera techniques we presented in this paper. Lastly, we present user-defined strategies that serve as design considerations for future smartphone AR applications utilizing a dual-camera setup. By conducting an exploratory study in this underexplored design space, we take a necessary first step toward uncovering new interaction possibilities. We invite future research to build upon our initial analysis and interaction concepts to create more immersive, expressive, and functional dual-camera AR experiences.








\begin{acks}
We thank our participants and the anonymous reviewers for their valuable feedback, which was instrumental in shaping this research.
\end{acks}


\begin{spacing}{1.0}  
\bibliographystyle{ACM-Reference-Format}
\bibliography{paper}


\begin{thebibliography}{50}


\ifx \showCODEN    \undefined \def \showCODEN     #1{\unskip}     \fi
\ifx \showDOI      \undefined \def \showDOI       #1{#1}\fi
\ifx \showISBNx    \undefined \def \showISBNx     #1{\unskip}     \fi
\ifx \showISBNxiii \undefined \def \showISBNxiii  #1{\unskip}     \fi
\ifx \showISSN     \undefined \def \showISSN      #1{\unskip}     \fi
\ifx \showLCCN     \undefined \def \showLCCN      #1{\unskip}     \fi
\ifx \shownote     \undefined \def \shownote      #1{#1}          \fi
\ifx \showarticletitle \undefined \def \showarticletitle #1{#1}   \fi
\ifx \showURL      \undefined \def \showURL       {\relax}        \fi
\providecommand\bibfield[2]{#2}
\providecommand\bibinfo[2]{#2}
\providecommand\natexlab[1]{#1}
\providecommand\showeprint[2][]{arXiv:#2}

\bibitem[\protect\citeauthoryear{Abtahi, Hough, Landay, and Follmer}{Abtahi et~al\mbox{.}}{2022}]%
        {abtahi2022beyond}
\bibfield{author}{\bibinfo{person}{Parastoo Abtahi}, \bibinfo{person}{Sidney~Q. Hough}, \bibinfo{person}{James~A. Landay}, {and} \bibinfo{person}{Sean Follmer}.} \bibinfo{year}{2022}\natexlab{}.
\newblock \showarticletitle{Beyond Being Real: A Sensorimotor Control Perspective on Interactions in Virtual Reality}. In \bibinfo{booktitle}{\emph{Proceedings of the 2022 CHI Conference on Human Factors in Computing Systems}} \emph{(\bibinfo{series}{CHI '22})}. Association for Computing Machinery, New York, NY, USA, \bibinfo{pages}{1--17}.
\newblock


\bibitem[\protect\citeauthoryear{Agrawal, Simon, Bech, B{\ae}rentsen, and Forchhammer}{Agrawal et~al\mbox{.}}{2019}]%
        {agrawal2019defining}
\bibfield{author}{\bibinfo{person}{Sarvesh Agrawal}, \bibinfo{person}{Ad{\`e}le Simon}, \bibinfo{person}{S{\o}ren Bech}, \bibinfo{person}{Klaus B{\ae}rentsen}, {and} \bibinfo{person}{S{\o}ren Forchhammer}.} \bibinfo{year}{2019}\natexlab{}.
\newblock \showarticletitle{Defining immersion: Literature review and implications for research on immersive audiovisual experiences}.
\newblock \bibinfo{journal}{\emph{Journal of Audio Engineering Society}} \bibinfo{volume}{68}, \bibinfo{number}{6} (\bibinfo{year}{2019}), \bibinfo{pages}{404--417}.
\newblock


\bibitem[\protect\citeauthoryear{Alem, Tecchia, and Huang}{Alem et~al\mbox{.}}{2011}]%
        {alem2011handsonvideo}
\bibfield{author}{\bibinfo{person}{Leila Alem}, \bibinfo{person}{Franco Tecchia}, {and} \bibinfo{person}{Weidong Huang}.} \bibinfo{year}{2011}\natexlab{}.
\newblock \showarticletitle{HandsOnVideo: Towards a Gesture based Mobile AR System for Remote Collaboration}.
\newblock In \bibinfo{booktitle}{\emph{Recent Trends of Mobile Collaborative Augmented Reality Systems}}. \bibinfo{publisher}{Springer}, \bibinfo{address}{New York, NY, USA}, \bibinfo{pages}{135--148}.
\newblock


\bibitem[\protect\citeauthoryear{Aliprantis, Konstantakis, Nikopoulou, Mylonas, and Caridakis}{Aliprantis et~al\mbox{.}}{2019}]%
        {aliprantis2019natural}
\bibfield{author}{\bibinfo{person}{John Aliprantis}, \bibinfo{person}{Markos Konstantakis}, \bibinfo{person}{Rozalia Nikopoulou}, \bibinfo{person}{Phivos Mylonas}, {and} \bibinfo{person}{George Caridakis}.} \bibinfo{year}{2019}\natexlab{}.
\newblock \showarticletitle{Natural Interaction in Augmented Reality Context.}. In \bibinfo{booktitle}{\emph{VIPERC@IRCDL}}. \bibinfo{pages}{50--61}.
\newblock


\bibitem[\protect\citeauthoryear{Anderson, Grossman, Matejka, and Fitzmaurice}{Anderson et~al\mbox{.}}{2013}]%
        {anderson2013youmove}
\bibfield{author}{\bibinfo{person}{Fraser Anderson}, \bibinfo{person}{Tovi Grossman}, \bibinfo{person}{Justin Matejka}, {and} \bibinfo{person}{George Fitzmaurice}.} \bibinfo{year}{2013}\natexlab{}.
\newblock \showarticletitle{YouMove: enhancing movement training with an augmented reality mirror}. In \bibinfo{booktitle}{\emph{Proceedings of the 26th Annual ACM Symposium on User Interface Software and Technology}} \emph{(\bibinfo{series}{UIST '13})}. Association for Computing Machinery, New York, NY, USA, \bibinfo{pages}{311–320}.
\newblock
\urldef\tempurl%
\url{https://doi.org/10.1145/2501988.2502045}
\showDOI{\tempurl}


\bibitem[\protect\citeauthoryear{Arora, Habib~Kazi, Grossman, Fitzmaurice, and Singh}{Arora et~al\mbox{.}}{2018}]%
        {arora2018symbiosissketch}
\bibfield{author}{\bibinfo{person}{Rahul Arora}, \bibinfo{person}{Rubaiat Habib~Kazi}, \bibinfo{person}{Tovi Grossman}, \bibinfo{person}{George Fitzmaurice}, {and} \bibinfo{person}{Karan Singh}.} \bibinfo{year}{2018}\natexlab{}.
\newblock \showarticletitle{SymbiosisSketch: Combining 2D \& 3D Sketching for Designing Detailed 3D Objects in Situ}. In \bibinfo{booktitle}{\emph{Proceedings of the 2018 CHI Conference on Human Factors in Computing Systems}} (Montreal QC, Canada) \emph{(\bibinfo{series}{CHI '18})}. Association for Computing Machinery, New York, NY, USA, \bibinfo{pages}{1–15}.
\newblock
\showISBNx{9781450356206}
\urldef\tempurl%
\url{https://doi.org/10.1145/3173574.3173759}
\showDOI{\tempurl}


\bibitem[\protect\citeauthoryear{Auda, Faltaous, Gruenefeld, Mayer, and Schneegass}{Auda et~al\mbox{.}}{2023}]%
        {auda2023actuality}
\bibfield{author}{\bibinfo{person}{Jonas Auda}, \bibinfo{person}{Sarah Faltaous}, \bibinfo{person}{Uwe Gruenefeld}, \bibinfo{person}{Sven Mayer}, {and} \bibinfo{person}{Stefan Schneegass}.} \bibinfo{year}{2023}\natexlab{}.
\newblock \showarticletitle{The Actuality-Time Continuum: Visualizing Interactions and Transitions Taking Place in Cross-Reality Systems}. In \bibinfo{booktitle}{\emph{2023 IEEE International Symposium on Mixed and Augmented Reality Adjunct (ISMAR-Adjunct)}}. IEEE, \bibinfo{pages}{35--40}.
\newblock


\bibitem[\protect\citeauthoryear{Babic, Perteneder, Reiterer, and Haller}{Babic et~al\mbox{.}}{2020}]%
        {babic2020simo}
\bibfield{author}{\bibinfo{person}{Teo Babic}, \bibinfo{person}{Florian Perteneder}, \bibinfo{person}{Harald Reiterer}, {and} \bibinfo{person}{Michael Haller}.} \bibinfo{year}{2020}\natexlab{}.
\newblock \showarticletitle{Simo: Interactions with distant displays by smartphones with simultaneous face and world tracking}. In \bibinfo{booktitle}{\emph{Extended Abstracts of the 2020 CHI Conference on Human Factors in Computing Systems}}. \bibinfo{pages}{1--12}.
\newblock


\bibitem[\protect\citeauthoryear{Bennett, Metatla, Roudaut, and Mekler}{Bennett et~al\mbox{.}}{2023}]%
        {bennett2023does}
\bibfield{author}{\bibinfo{person}{Dan Bennett}, \bibinfo{person}{Oussama Metatla}, \bibinfo{person}{Anne Roudaut}, {and} \bibinfo{person}{Elisa~D. Mekler}.} \bibinfo{year}{2023}\natexlab{}.
\newblock \showarticletitle{How does HCI Understand Human Agency and Autonomy?}. In \bibinfo{booktitle}{\emph{Proceedings of the 2023 CHI Conference on Human Factors in Computing Systems}} (Hamburg, Germany) \emph{(\bibinfo{series}{CHI '23})}. Association for Computing Machinery, New York, NY, USA, Article \bibinfo{articleno}{375}, \bibinfo{numpages}{18}~pages.
\newblock
\showISBNx{9781450394215}
\urldef\tempurl%
\url{https://doi.org/10.1145/3544548.3580651}
\showDOI{\tempurl}


\bibitem[\protect\citeauthoryear{Brasier, Pietriga, and Appert}{Brasier et~al\mbox{.}}{2021}]%
        {brasier2021ar}
\bibfield{author}{\bibinfo{person}{Eugenie Brasier}, \bibinfo{person}{Emmanuel Pietriga}, {and} \bibinfo{person}{Caroline Appert}.} \bibinfo{year}{2021}\natexlab{}.
\newblock \showarticletitle{AR-enhanced Widgets for Smartphone-centric Interaction}. In \bibinfo{booktitle}{\emph{Proceedings of the 23rd International Conference on Mobile Human-Computer Interaction}}. \bibinfo{pages}{1--12}.
\newblock


\bibitem[\protect\citeauthoryear{Carrol}{Carrol}{1999}]%
        {carrol1999five}
\bibfield{author}{\bibinfo{person}{John~M Carrol}.} \bibinfo{year}{1999}\natexlab{}.
\newblock \showarticletitle{Five reasons for scenario-based design}. In \bibinfo{booktitle}{\emph{Proceedings of the 32nd annual hawaii international conference on systems sciences. 1999. hicss-32. abstracts and cd-rom of full papers}}. IEEE, \bibinfo{pages}{11--pp}.
\newblock


\bibitem[\protect\citeauthoryear{Carroll}{Carroll}{2003}]%
        {carroll2003making}
\bibfield{author}{\bibinfo{person}{John~M Carroll}.} \bibinfo{year}{2003}\natexlab{}.
\newblock \bibinfo{booktitle}{\emph{Making use: scenario-based design of human-computer interactions}}.
\newblock \bibinfo{publisher}{MIT press}.
\newblock


\bibitem[\protect\citeauthoryear{Chudinov}{Chudinov}{2014}]%
        {chudinov2014approximate}
\bibfield{author}{\bibinfo{person}{Peter Chudinov}.} \bibinfo{year}{2014}\natexlab{}.
\newblock \showarticletitle{Approximate analytical description of the projectile motion with a quadratic drag force}.
\newblock \bibinfo{journal}{\emph{Athens J. Nat. Formal Sci}}  \bibinfo{volume}{1} (\bibinfo{year}{2014}), \bibinfo{pages}{97--106}.
\newblock


\bibitem[\protect\citeauthoryear{Endsley, Sprehn, Brill, Ryan, Vincent, and Martin}{Endsley et~al\mbox{.}}{2017}]%
        {endsley2017augmented}
\bibfield{author}{\bibinfo{person}{Tristan~C Endsley}, \bibinfo{person}{Kelly~A Sprehn}, \bibinfo{person}{Ryan~M Brill}, \bibinfo{person}{Kimberly~J Ryan}, \bibinfo{person}{Emily~C Vincent}, {and} \bibinfo{person}{James~M Martin}.} \bibinfo{year}{2017}\natexlab{}.
\newblock \showarticletitle{Augmented reality design heuristics: Designing for dynamic interactions}. In \bibinfo{booktitle}{\emph{Proceedings of the human factors and ergonomics society annual meeting}}, Vol.~\bibinfo{volume}{61}. Sage Publications Sage CA: Los Angeles, CA, \bibinfo{pages}{2100--2104}.
\newblock


\bibitem[\protect\citeauthoryear{Feuchtner and M{\"u}ller}{Feuchtner and M{\"u}ller}{2017}]%
        {feuchtner2017extending}
\bibfield{author}{\bibinfo{person}{Tiare Feuchtner} {and} \bibinfo{person}{J{\"o}rg M{\"u}ller}.} \bibinfo{year}{2017}\natexlab{}.
\newblock \showarticletitle{Extending the body for interaction with reality}. In \bibinfo{booktitle}{\emph{Proceedings of the 2017 CHI Conference on Human Factors in Computing Systems}}. \bibinfo{pages}{5145--5157}.
\newblock


\bibitem[\protect\citeauthoryear{Genay, L{\'e}cuyer, and Hachet}{Genay et~al\mbox{.}}{2021}]%
        {genay2021being}
\bibfield{author}{\bibinfo{person}{Ad{\'e}la{\"\i}de Genay}, \bibinfo{person}{Anatole L{\'e}cuyer}, {and} \bibinfo{person}{Martin Hachet}.} \bibinfo{year}{2021}\natexlab{}.
\newblock \showarticletitle{Being an avatar “for real”: a survey on virtual embodiment in augmented reality}.
\newblock \bibinfo{journal}{\emph{IEEE Transactions on Visualization and Computer Graphics}} \bibinfo{volume}{28}, \bibinfo{number}{12} (\bibinfo{year}{2021}), \bibinfo{pages}{5071--5090}.
\newblock


\bibitem[\protect\citeauthoryear{Grubert, Ofek, Pahud, Kranz, and Schmalstieg}{Grubert et~al\mbox{.}}{2016}]%
        {grubert2016glasshands}
\bibfield{author}{\bibinfo{person}{Jens Grubert}, \bibinfo{person}{Eyal Ofek}, \bibinfo{person}{Michel Pahud}, \bibinfo{person}{Matthias Kranz}, {and} \bibinfo{person}{Dieter Schmalstieg}.} \bibinfo{year}{2016}\natexlab{}.
\newblock \showarticletitle{Glasshands: Interaction around unmodified mobile devices using sunglasses}. In \bibinfo{booktitle}{\emph{Proceedings of the 2016 ACM International Conference on Interactive Surfaces and Spaces}}. \bibinfo{pages}{215--224}.
\newblock


\bibitem[\protect\citeauthoryear{Guest, Bunce, and Johnson}{Guest et~al\mbox{.}}{2006}]%
        {guest2006many}
\bibfield{author}{\bibinfo{person}{Greg Guest}, \bibinfo{person}{Arwen Bunce}, {and} \bibinfo{person}{Laura Johnson}.} \bibinfo{year}{2006}\natexlab{}.
\newblock \showarticletitle{How many interviews are enough? An experiment with data saturation and variability}.
\newblock \bibinfo{journal}{\emph{Field methods}} \bibinfo{volume}{18}, \bibinfo{number}{1} (\bibinfo{year}{2006}), \bibinfo{pages}{59--82}.
\newblock


\bibitem[\protect\citeauthoryear{Iftikhar, Haq, Younus, Sardar, Arif, Javed, and Shahid}{Iftikhar et~al\mbox{.}}{2021}]%
        {iftikhar2021designing}
\bibfield{author}{\bibinfo{person}{Zainab Iftikhar}, \bibinfo{person}{Qutaiba Rohan~ul Haq}, \bibinfo{person}{Osama Younus}, \bibinfo{person}{Taha Sardar}, \bibinfo{person}{Hammad Arif}, \bibinfo{person}{Mobin Javed}, {and} \bibinfo{person}{Suleman Shahid}.} \bibinfo{year}{2021}\natexlab{}.
\newblock \showarticletitle{Designing parental monitoring and control technology: A systematic review}. In \bibinfo{booktitle}{\emph{Human-Computer Interaction--INTERACT 2021: 18th IFIP TC 13 International Conference, Bari, Italy, August 30--September 3, 2021, Proceedings, Part IV 18}}. Springer, \bibinfo{pages}{676--700}.
\newblock


\bibitem[\protect\citeauthoryear{Iftikhar, Ma, and Huang}{Iftikhar et~al\mbox{.}}{2023}]%
        {iftikhar2023together}
\bibfield{author}{\bibinfo{person}{Zainab Iftikhar}, \bibinfo{person}{Yumeng Ma}, {and} \bibinfo{person}{Jeff Huang}.} \bibinfo{year}{2023}\natexlab{}.
\newblock \showarticletitle{“Together but not together”: Evaluating Typing Indicators for Interaction-Rich Communication}. In \bibinfo{booktitle}{\emph{Proceedings of the 2023 CHI Conference on Human Factors in Computing Systems}}. \bibinfo{pages}{1--12}.
\newblock


\bibitem[\protect\citeauthoryear{Inc.}{Inc.}{2024}]%
        {apple2024arkitrealitykit}
\bibfield{author}{\bibinfo{person}{Apple Inc.}} \bibinfo{year}{2024}\natexlab{}.
\newblock \bibinfo{title}{ARKit and RealityKit}.
\newblock \bibinfo{howpublished}{\url{https://developer.apple.com/augmented-reality/}}.
\newblock


\bibitem[\protect\citeauthoryear{Kim and Lee}{Kim and Lee}{2016}]%
        {kim2016touch}
\bibfield{author}{\bibinfo{person}{Minseok Kim} {and} \bibinfo{person}{Jae~Yeol Lee}.} \bibinfo{year}{2016}\natexlab{}.
\newblock \showarticletitle{Touch and hand gesture-based interactions for directly manipulating 3D virtual objects in mobile augmented reality}.
\newblock \bibinfo{journal}{\emph{Multimedia Tools and Applications}}  \bibinfo{volume}{75} (\bibinfo{year}{2016}), \bibinfo{pages}{16529--16550}.
\newblock


\bibitem[\protect\citeauthoryear{Latoschik, Lugrin, and Roth}{Latoschik et~al\mbox{.}}{2016}]%
        {latoschik2016fakemi}
\bibfield{author}{\bibinfo{person}{Marc~Erich Latoschik}, \bibinfo{person}{Jean-Luc Lugrin}, {and} \bibinfo{person}{Daniel Roth}.} \bibinfo{year}{2016}\natexlab{}.
\newblock \showarticletitle{FakeMi: A fake mirror system for avatar embodiment studies}. In \bibinfo{booktitle}{\emph{Proceedings of the 22nd ACM Conference on Virtual Reality Software and Technology}}. \bibinfo{pages}{73--76}.
\newblock


\bibitem[\protect\citeauthoryear{Li, Zhang, Liu, Yang, Fu, Tian, Han, and Fan}{Li et~al\mbox{.}}{2021}]%
        {li2021vmirror}
\bibfield{author}{\bibinfo{person}{Nianlong Li}, \bibinfo{person}{Zhengquan Zhang}, \bibinfo{person}{Can Liu}, \bibinfo{person}{Zengyao Yang}, \bibinfo{person}{Yinan Fu}, \bibinfo{person}{Feng Tian}, \bibinfo{person}{Teng Han}, {and} \bibinfo{person}{Mingming Fan}.} \bibinfo{year}{2021}\natexlab{}.
\newblock \showarticletitle{Vmirror: Enhancing the interaction with occluded or distant objects in vr with virtual mirrors}. In \bibinfo{booktitle}{\emph{Proceedings of the 2021 CHI Conference on Human Factors in Computing Systems}}. \bibinfo{pages}{1--11}.
\newblock


\bibitem[\protect\citeauthoryear{Loorak, Zhou, Trinh, Zhao, and Li}{Loorak et~al\mbox{.}}{2019}]%
        {loorak2019hand}
\bibfield{author}{\bibinfo{person}{Mona~Hosseinkhani Loorak}, \bibinfo{person}{Wei Zhou}, \bibinfo{person}{Ha Trinh}, \bibinfo{person}{Jian Zhao}, {and} \bibinfo{person}{Wei Li}.} \bibinfo{year}{2019}\natexlab{}.
\newblock \showarticletitle{Hand-over-face input sensing for interaction with smartphones through the built-in camera}. In \bibinfo{booktitle}{\emph{Proceedings of the 21st International Conference on Human-Computer Interaction with Mobile Devices and Services}}. \bibinfo{pages}{1--12}.
\newblock


\bibitem[\protect\citeauthoryear{Ma, Qian, Zhou, and Huang}{Ma et~al\mbox{.}}{2023}]%
        {ma2023focalpoint}
\bibfield{author}{\bibinfo{person}{Jiaju Ma}, \bibinfo{person}{Jing Qian}, \bibinfo{person}{Tongyu Zhou}, {and} \bibinfo{person}{Jeff Huang}.} \bibinfo{year}{2023}\natexlab{}.
\newblock \showarticletitle{FocalPoint: Adaptive Direct Manipulation for Selecting Small 3D Virtual Objects}.
\newblock \bibinfo{journal}{\emph{Proceedings of the ACM on Interactive, Mobile, Wearable and Ubiquitous Technologies}} \bibinfo{volume}{7}, \bibinfo{number}{1} (\bibinfo{year}{2023}), \bibinfo{pages}{1--26}.
\newblock


\bibitem[\protect\citeauthoryear{Mayer, Chiossi, and Mayer}{Mayer et~al\mbox{.}}{2024}]%
        {mayer2024crossing}
\bibfield{author}{\bibinfo{person}{Elisabeth Mayer}, \bibinfo{person}{Francesco Chiossi}, {and} \bibinfo{person}{Sven Mayer}.} \bibinfo{year}{2024}\natexlab{}.
\newblock \showarticletitle{Crossing Mixed Realities: A Review for Transitional Interfaces Design}.
\newblock \bibinfo{journal}{\emph{Proceedings of Mensch und Computer 2024}} (\bibinfo{year}{2024}), \bibinfo{pages}{629--634}.
\newblock


\bibitem[\protect\citeauthoryear{Moln{\'a}r and Sz{\H{u}}ts}{Moln{\'a}r and Sz{\H{u}}ts}{2019}]%
        {molnar2019augmented}
\bibfield{author}{\bibinfo{person}{Gy{\"o}rgy Moln{\'a}r} {and} \bibinfo{person}{Zolt{\'a}n Sz{\H{u}}ts}.} \bibinfo{year}{2019}\natexlab{}.
\newblock \showarticletitle{Augmented reality, games and art: immersion and flow}.
\newblock \bibinfo{journal}{\emph{Augmented Reality Games I: Understanding the Pok{\'e}mon GO Phenomenon}} (\bibinfo{year}{2019}), \bibinfo{pages}{61--67}.
\newblock


\bibitem[\protect\citeauthoryear{Nagai, Fujita, Takashima, and Kitamura}{Nagai et~al\mbox{.}}{2022}]%
        {nagai2022handygaze}
\bibfield{author}{\bibinfo{person}{Takahiro Nagai}, \bibinfo{person}{Kazuyuki Fujita}, \bibinfo{person}{Kazuki Takashima}, {and} \bibinfo{person}{Yoshifumi Kitamura}.} \bibinfo{year}{2022}\natexlab{}.
\newblock \showarticletitle{HandyGaze: A Gaze Tracking Technique for Room-Scale Environments using a Single Smartphone}.
\newblock \bibinfo{journal}{\emph{Proceedings of the ACM on Human-Computer Interaction}} \bibinfo{volume}{6}, \bibinfo{number}{ISS} (\bibinfo{year}{2022}), \bibinfo{pages}{143--160}.
\newblock


\bibitem[\protect\citeauthoryear{Niantic}{Niantic}{2024}]%
        {niantic2024pokemongo}
\bibfield{author}{\bibinfo{person}{Inc. Niantic}.} \bibinfo{year}{2024}\natexlab{}.
\newblock \bibinfo{title}{Pokémon GO}.
\newblock \bibinfo{howpublished}{\url{https://pokemongolive.com/}}.
\newblock


\bibitem[\protect\citeauthoryear{O’Hare, Dekoninck, Mombeshora, Martens, Becattini, and Boujut}{O’Hare et~al\mbox{.}}{2020}]%
        {o2020defining}
\bibfield{author}{\bibinfo{person}{Jamie O’Hare}, \bibinfo{person}{Elies Dekoninck}, \bibinfo{person}{Mendy Mombeshora}, \bibinfo{person}{Philippe Martens}, \bibinfo{person}{Niccol{\`o} Becattini}, {and} \bibinfo{person}{Jean-Francois Boujut}.} \bibinfo{year}{2020}\natexlab{}.
\newblock \showarticletitle{Defining requirements for an Augmented Reality system to overcome the challenges of creating and using design representations in co-design sessions}.
\newblock \bibinfo{journal}{\emph{CoDesign}} \bibinfo{volume}{16}, \bibinfo{number}{2} (\bibinfo{year}{2020}), \bibinfo{pages}{111--134}.
\newblock


\bibitem[\protect\citeauthoryear{Pointecker, Friedl, Schwajda, Jetter, and Anthes}{Pointecker et~al\mbox{.}}{2022}]%
        {pointecker2022bridging}
\bibfield{author}{\bibinfo{person}{Fabian Pointecker}, \bibinfo{person}{Judith Friedl}, \bibinfo{person}{Daniel Schwajda}, \bibinfo{person}{Hans-Christian Jetter}, {and} \bibinfo{person}{Christoph Anthes}.} \bibinfo{year}{2022}\natexlab{}.
\newblock \showarticletitle{Bridging the gap across realities: Visual transitions between virtual and augmented reality}. In \bibinfo{booktitle}{\emph{2022 IEEE international symposium on mixed and augmented reality (ISMAR)}}. IEEE, \bibinfo{pages}{827--836}.
\newblock


\bibitem[\protect\citeauthoryear{Qian, Ma, Li, Attal, Lai, Tompkin, Hughes, and Huang}{Qian et~al\mbox{.}}{2019}]%
        {qian2019portal}
\bibfield{author}{\bibinfo{person}{Jing Qian}, \bibinfo{person}{Jiaju Ma}, \bibinfo{person}{Xiangyu Li}, \bibinfo{person}{Benjamin Attal}, \bibinfo{person}{Haoming Lai}, \bibinfo{person}{James Tompkin}, \bibinfo{person}{John~F Hughes}, {and} \bibinfo{person}{Jeff Huang}.} \bibinfo{year}{2019}\natexlab{}.
\newblock \showarticletitle{Portal-ble: Intuitive free-hand manipulation in unbounded smartphone-based augmented reality}. In \bibinfo{booktitle}{\emph{Proceedings of the 32nd Annual ACM Symposium on User Interface Software and Technology}}. \bibinfo{pages}{133--145}.
\newblock


\bibitem[\protect\citeauthoryear{Qian, Shamma, Avrahami, and Biehl}{Qian et~al\mbox{.}}{2020a}]%
        {qian2020modality}
\bibfield{author}{\bibinfo{person}{Jing Qian}, \bibinfo{person}{David~A. Shamma}, \bibinfo{person}{Daniel Avrahami}, {and} \bibinfo{person}{Jacob Biehl}.} \bibinfo{year}{2020}\natexlab{a}.
\newblock \showarticletitle{Modality and Depth in Touchless Smartphone Augmented Reality Interactions}. In \bibinfo{booktitle}{\emph{Proceedings of the 2020 ACM International Conference on Interactive Media Experiences}} (Cornella, Barcelona, Spain) \emph{(\bibinfo{series}{IMX '20})}. Association for Computing Machinery, New York, NY, USA, \bibinfo{pages}{74–81}.
\newblock
\showISBNx{9781450379762}
\urldef\tempurl%
\url{https://doi.org/10.1145/3391614.3393648}
\showDOI{\tempurl}


\bibitem[\protect\citeauthoryear{Qian, Young-Ng, Li, Cheung, Yang, and Huang}{Qian et~al\mbox{.}}{2020b}]%
        {qian2020portalware}
\bibfield{author}{\bibinfo{person}{Jing Qian}, \bibinfo{person}{Meredith Young-Ng}, \bibinfo{person}{Xiangyu Li}, \bibinfo{person}{Angel Cheung}, \bibinfo{person}{Fumeng Yang}, {and} \bibinfo{person}{Jeff Huang}.} \bibinfo{year}{2020}\natexlab{b}.
\newblock \showarticletitle{Portalware: A Smartphone-Wearable Dual-Display System for Expanding the Free-Hand Interaction Region in Augmented Reality}. In \bibinfo{booktitle}{\emph{Extended Abstracts of the 2020 CHI Conference on Human Factors in Computing Systems}} \emph{(\bibinfo{series}{CHI EA '20})}. Association for Computing Machinery, New York, NY, USA, \bibinfo{pages}{1–8}.
\newblock
\showISBNx{9781450368193}
\urldef\tempurl%
\url{https://doi.org/10.1145/3334480.3383079}
\showDOI{\tempurl}


\bibitem[\protect\citeauthoryear{Roo and Hachet}{Roo and Hachet}{2017}]%
        {roo2017one}
\bibfield{author}{\bibinfo{person}{Joan~Sol Roo} {and} \bibinfo{person}{Martin Hachet}.} \bibinfo{year}{2017}\natexlab{}.
\newblock \showarticletitle{One reality: Augmenting how the physical world is experienced by combining multiple mixed reality modalities}. In \bibinfo{booktitle}{\emph{Proceedings of the 30th annual ACM symposium on user interface software and technology}}. \bibinfo{pages}{787--795}.
\newblock


\bibitem[\protect\citeauthoryear{Sakashita, Kim, Woodard, Zhang, and Guimbreti{\`e}re}{Sakashita et~al\mbox{.}}{2023}]%
        {sakashita2023vroxy}
\bibfield{author}{\bibinfo{person}{Mose Sakashita}, \bibinfo{person}{Hyunju Kim}, \bibinfo{person}{Brandon Woodard}, \bibinfo{person}{Ruidong Zhang}, {and} \bibinfo{person}{Fran{\c{c}}ois Guimbreti{\`e}re}.} \bibinfo{year}{2023}\natexlab{}.
\newblock \showarticletitle{VRoxy: Wide-Area Collaboration From an Office Using a VR-Driven Robotic Proxy}. In \bibinfo{booktitle}{\emph{Proceedings of the 36th Annual ACM Symposium on User Interface Software and Technology}}. \bibinfo{pages}{1--13}.
\newblock


\bibitem[\protect\citeauthoryear{Shan, Sun, Tart, Woodard, Humer, and Eckhardt}{Shan et~al\mbox{.}}{2024}]%
        {shan2024work}
\bibfield{author}{\bibinfo{person}{Kaili Shan}, \bibinfo{person}{Tiger Sun}, \bibinfo{person}{Jarrod Tart}, \bibinfo{person}{Brandon Woodard}, \bibinfo{person}{Irene Humer}, {and} \bibinfo{person}{Christian Eckhardt}.} \bibinfo{year}{2024}\natexlab{}.
\newblock \showarticletitle{Work-in-Progress—Virtual Learning Laboratories for High School Chemistry Lab: An Immersive Learning User Study}.
\newblock \bibinfo{journal}{\emph{Immersive Learning Research-Academic}} (\bibinfo{year}{2024}), \bibinfo{pages}{61--71}.
\newblock


\bibitem[\protect\citeauthoryear{Snap}{Snap}{2024}]%
        {snap2024snapchat}
\bibfield{author}{\bibinfo{person}{Inc. Snap}.} \bibinfo{year}{2024}\natexlab{}.
\newblock \bibinfo{title}{Snapchat AR Filters}.
\newblock \bibinfo{howpublished}{\url{https://whatis.snapchat.com/}}.
\newblock


\bibitem[\protect\citeauthoryear{Studios}{Studios}{2024}]%
        {mojang2024minecraft}
\bibfield{author}{\bibinfo{person}{Mojang Studios}.} \bibinfo{year}{2024}\natexlab{}.
\newblock \bibinfo{title}{Minecraft}.
\newblock \bibinfo{howpublished}{\url{https://www.minecraft.net/}}.
\newblock


\bibitem[\protect\citeauthoryear{Surale, Gupta, Hancock, and Vogel}{Surale et~al\mbox{.}}{2019}]%
        {surale2019tabletinvr}
\bibfield{author}{\bibinfo{person}{Hemant~Bhaskar Surale}, \bibinfo{person}{Aakar Gupta}, \bibinfo{person}{Mark Hancock}, {and} \bibinfo{person}{Daniel Vogel}.} \bibinfo{year}{2019}\natexlab{}.
\newblock \showarticletitle{Tabletinvr: Exploring the design space for using a multi-touch tablet in virtual reality}. In \bibinfo{booktitle}{\emph{Proceedings of the 2019 CHI Conference on Human Factors in Computing Systems}}. \bibinfo{pages}{1--13}.
\newblock


\bibitem[\protect\citeauthoryear{Tseng and Tseng}{Tseng and Tseng}{2013}]%
        {tseng2013cohesion}
\bibfield{author}{\bibinfo{person}{Chiao-I Tseng} {and} \bibinfo{person}{Chiao-I Tseng}.} \bibinfo{year}{2013}\natexlab{}.
\newblock \bibinfo{booktitle}{\emph{Cohesion in film}}.
\newblock \bibinfo{publisher}{Springer}.
\newblock


\bibitem[\protect\citeauthoryear{Tyack and Wyeth}{Tyack and Wyeth}{2017}]%
        {tyack2017exploring}
\bibfield{author}{\bibinfo{person}{April Tyack} {and} \bibinfo{person}{Peta Wyeth}.} \bibinfo{year}{2017}\natexlab{}.
\newblock \showarticletitle{Exploring relatedness in single-player video game play}. In \bibinfo{booktitle}{\emph{Proceedings of the 29th Australian conference on computer-human interaction}}. \bibinfo{pages}{422--427}.
\newblock


\bibitem[\protect\citeauthoryear{Vanukuru, Weng, Ranjan, Hopkins, Banic, Gross, and Do}{Vanukuru et~al\mbox{.}}{2023}]%
        {vanukuru2023dualstream}
\bibfield{author}{\bibinfo{person}{Rishi Vanukuru}, \bibinfo{person}{Suibi Che-Chuan Weng}, \bibinfo{person}{Krithik Ranjan}, \bibinfo{person}{Torin Hopkins}, \bibinfo{person}{Amy Banic}, \bibinfo{person}{Mark~D Gross}, {and} \bibinfo{person}{Ellen Yi-Luen Do}.} \bibinfo{year}{2023}\natexlab{}.
\newblock \showarticletitle{DualStream: Spatially Sharing Selves and Surroundings using Mobile Devices and Augmented Reality}. In \bibinfo{booktitle}{\emph{2023 IEEE International Symposium on Mixed and Augmented Reality (ISMAR)}}. IEEE, \bibinfo{pages}{138--147}.
\newblock


\bibitem[\protect\citeauthoryear{Weber, Weibel, and Mast}{Weber et~al\mbox{.}}{2021}]%
        {weber2021get}
\bibfield{author}{\bibinfo{person}{Stefan Weber}, \bibinfo{person}{David Weibel}, {and} \bibinfo{person}{Fred~W Mast}.} \bibinfo{year}{2021}\natexlab{}.
\newblock \showarticletitle{How to get there when you are there already? Defining presence in virtual reality and the importance of perceived realism}.
\newblock \bibinfo{journal}{\emph{Frontiers in psychology}}  \bibinfo{volume}{12} (\bibinfo{year}{2021}), \bibinfo{pages}{628298}.
\newblock


\bibitem[\protect\citeauthoryear{Wu and Li}{Wu and Li}{2024}]%
        {wu2024enable}
\bibfield{author}{\bibinfo{person}{Qinyang Wu} {and} \bibinfo{person}{Chen Li}.} \bibinfo{year}{2024}\natexlab{}.
\newblock \showarticletitle{Enable Natural User Interactions in Handheld Mobile Augmented Reality through Image Computing}. In \bibinfo{booktitle}{\emph{Proceedings of the 2024 ACM Symposium on Spatial User Interaction}}. \bibinfo{pages}{1--2}.
\newblock


\bibitem[\protect\citeauthoryear{Xu, Barba, Radu, Gandy, Shemaka, Schrank, MacIntyre, and Tseng}{Xu et~al\mbox{.}}{2011}]%
        {xu2011pre}
\bibfield{author}{\bibinfo{person}{Yan Xu}, \bibinfo{person}{Evan Barba}, \bibinfo{person}{Iulian Radu}, \bibinfo{person}{Maribeth Gandy}, \bibinfo{person}{Richard Shemaka}, \bibinfo{person}{Brian Schrank}, \bibinfo{person}{Blair MacIntyre}, {and} \bibinfo{person}{Tony Tseng}.} \bibinfo{year}{2011}\natexlab{}.
\newblock \showarticletitle{Pre-patterns for designing embodied interactions in handheld augmented reality games}. In \bibinfo{booktitle}{\emph{2011 IEEE International Symposium on Mixed and Augmented Reality-Arts, Media, and Humanities}}. IEEE, \bibinfo{pages}{19--28}.
\newblock


\bibitem[\protect\citeauthoryear{Yeo, Wu, Kim, Lee, Kim, Oh, Takagi, Woo, Koike, and Quigley}{Yeo et~al\mbox{.}}{2023}]%
        {yeo2023omnisense}
\bibfield{author}{\bibinfo{person}{Hui-Shyong Yeo}, \bibinfo{person}{Erwin Wu}, \bibinfo{person}{Daehwa Kim}, \bibinfo{person}{Juyoung Lee}, \bibinfo{person}{Hyung-il Kim}, \bibinfo{person}{Seo~Young Oh}, \bibinfo{person}{Luna Takagi}, \bibinfo{person}{Woontack Woo}, \bibinfo{person}{Hideki Koike}, {and} \bibinfo{person}{Aaron~John Quigley}.} \bibinfo{year}{2023}\natexlab{}.
\newblock \showarticletitle{OmniSense: Exploring Novel Input Sensing and Interaction Techniques on Mobile Device with an Omni-Directional Camera}. In \bibinfo{booktitle}{\emph{Proceedings of the 2023 CHI Conference on Human Factors in Computing Systems}}. \bibinfo{pages}{1--18}.
\newblock


\bibitem[\protect\citeauthoryear{Zhao, Fanello, and Guo}{Zhao et~al\mbox{.}}{2023}]%
        {zhao2023multi}
\bibfield{author}{\bibinfo{person}{Yiqin Zhao}, \bibinfo{person}{Sean Fanello}, {and} \bibinfo{person}{Tian Guo}.} \bibinfo{year}{2023}\natexlab{}.
\newblock \showarticletitle{Multi-camera lighting estimation for photorealistic front-facing mobile augmented reality}. In \bibinfo{booktitle}{\emph{Proceedings of the 24th International Workshop on Mobile Computing Systems and Applications}}. \bibinfo{pages}{68--73}.
\newblock


\bibitem[\protect\citeauthoryear{Zhu and Grossman}{Zhu and Grossman}{2020}]%
        {zhu2020bishare}
\bibfield{author}{\bibinfo{person}{Fengyuan Zhu} {and} \bibinfo{person}{Tovi Grossman}.} \bibinfo{year}{2020}\natexlab{}.
\newblock \showarticletitle{Bishare: Exploring bidirectional interactions between smartphones and head-mounted augmented reality}. In \bibinfo{booktitle}{\emph{Proceedings of the 2020 CHI Conference on Human Factors in Computing Systems}}. \bibinfo{pages}{1--14}.
\newblock


\end{thebibliography}
\end{spacing}

\clearpage
\section{Appendix}
\appendix
\section{Application Scenarios}

\begin{enumerate}
    \item Architecture \& Object Arrangement (P18, P30): Using hand-to-camera gestures to position and manipulate virtual furniture or building elements in real space.
    \item Gesture-Based Page Flipping (P22): Winking or other face movements to turn pages in e-books, social media feeds, or document viewers.
    \item Virtual Museum Augmentation (P21): Closing one eye to switch between different versions or layers of an artwork, deepening engagement and exploring embodiment with art pieces.
    \item Character Interaction via Facial Gestures (P19, P23): Animating or commanding virtual characters or pets through smiles, winks, or other facial inputs.
    \item Sports Training \& Aiming (P9, P18, P19, P25): Using precise eye- or face-based gestures to aim virtual projectiles and launch virtual projectiles.
    \item 
\end{enumerate}
\begin{figure}
    \centering
    \includegraphics[width=0.7\linewidth]{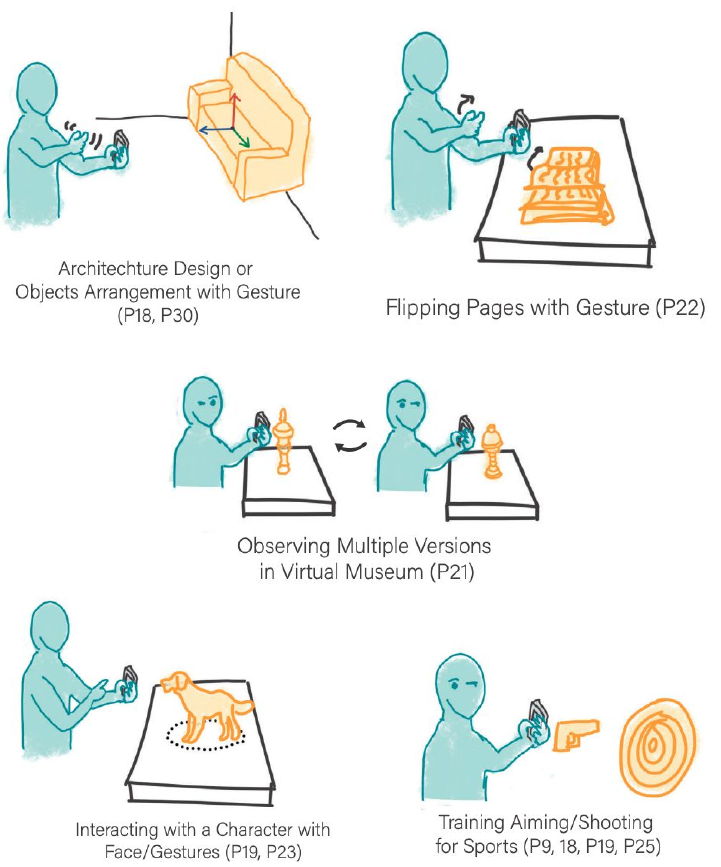}
    \caption{After trying out \textit{Cam-2-Cam} interactions, participants shared application scenarios that they thought could greatly leverage a dual-camera setup to incorporate physical gestures.}
    \label{fig:future-app}
\end{figure}

These sketches paired with our qualitative findings underscore how dual-camera AR could support novel interaction methods across art, education, gaming, training, and design, and motivate a host of future empirical studies.

\end{document}